\documentclass[%
 reprint,
 amsmath,amssymb,
 aps,
 prx,
]{revtex4-2}
\usepackage{physics} %added by author!
\usepackage{graphicx}% Include figure files
\usepackage{dcolumn}% Align table columns on decimal point
\usepackage{bm}% bold math
\usepackage{lineno}%{ \linenumbersep}{5pt}
\usepackage{xcolor}
\usepackage{siunitx}
\makeatletter
\providecommand \@ifxundefined [1]{%
 \@ifx{#1\undefined}
}%
\providecommand \@ifnum [1]{%
 \ifnum #1\expandafter \@firstoftwo
 \else \expandafter \@secondoftwo
 \fi
}%
\providecommand \@ifx [1]{%
 \ifx #1\expandafter \@firstoftwo
 \else \expandafter \@secondoftwo
 \fi
}%
\providecommand \natexlab [1]{#1}%
\providecommand \bibnamefont  [1]{#1}%
\providecommand \bibfnamefont [1]{#1}%
\providecommand \citenamefont [1]{#1}%
\providecommand \href@noop [0]{\@secondoftwo}%
\providecommand \href [0]{\begingroup \@sanitize@url \@href}%
\providecommand \@href[1]{\@@startlink{#1}\@@href}%
\providecommand \@@href[1]{\endgroup#1\@@endlink}%
\providecommand \@sanitize@url [0]{\catcode `\\12\catcode `\$12\catcode
  `\&12\catcode `\#12\catcode `\^12\catcode `\_12\catcode `\%12\relax}%
\providecommand \@@startlink[1]{}%
\providecommand \@@endlink[0]{}%
\providecommand \url  [0]{\begingroup\@sanitize@url \@url }%
\providecommand \@url [1]{\endgroup\@href {#1}{\urlprefix }}%
\providecommand \urlprefix  [0]{URL }%
\providecommand \Eprint [0]{\href }%
\providecommand \selectlanguage [0]{\@gobble}%
\providecommand \bibinfo  [0]{\@secondoftwo}%
\providecommand \bibfield  [0]{\@secondoftwo}%
\providecommand \BibitemOpen [0]{}%
\providecommand \EOS [0]{\spacefactor3000\relax}%
\providecommand \BibitemShut  [1]{\csname bibitem#1\endcsname}%
\let\auto@bib@innerbib\@empty

\begin{document}
%\linenumbers
%EMAILS HAVE TO BE HERE!!!

\preprint{APS/123-QED}

\title{Discovery of ST2 centers in natural and CVD diamond}% Force line breaks with \\
\author{Jonas Foglszinger$^1$}
\author{Andrej Denisenko$^1$}%
\author{Georgy V. Astakhov$^2$}
\author{Lev Kazak$^3$}
\author{Petr Siyushev$^4$}
\author{Alexander M. Zaitsev$^5$}
\author{Roman Kolesov$^1$}%
\author{J\"org Wrachtrup$^{1,6}$}%
\affiliation{$^1$3rd Institute of Physics, University of Stuttgart,70569 Stuttgart, Germany}%
\affiliation{$^2$Helmholtz-Zentrum Dresden-Rossendorf,
Institute of Ion Beam Physics and Materials Research,
01328 Dresden, Germany}
\affiliation{$^3$University Institute for Quantum Optics, Ulm University, Albert-Einstein-Allee 11, D-89081 Ulm, Germany}
\affiliation{$^4$Institute for Materials Research, Hasselt University, Wetenschapspark 1, 3590 Diepenbeek, Belgium}
\affiliation{$^5$ College of Staten Island (CUNY), 2800 Victory Blvd., Staten Island, NY 10312, USA}
\affiliation{$^6$ Max Planck Institute for Solid State Research}
\date{\today}% It is always \today, today,
\begin{abstract}
The ST2 center is an optically addressable point defect in diamond that facilitates spin initialization and readout.
However, while this study presents the discovery of ST2 centers first observed in a natural diamond and provides a reliable technique  for artificially creating them, its chemical structure remains unknown.
To assess the potential of ST2, we map out its basic optical characteristics, reveal its electronic level structure, and quantify the intrinsic transition rates.
Furthermore, we investigate its response to microwaves, static magnetic fields, and the polarization of excitation laser light, revealing twelve inequivalent orientations of the ST2 center.
Simultaneous exposure to microwaves and static magnetic fields also reveals an exceptionally wide acceptance angle for sensing strong magnetic fields, unlike the well-established NV center, which is sensitive only within a narrow cone aligned with its symmetry axis.
This finding establishes the ST2 center as a highly promising candidate for nanoscale quantum sensing.
\vspace{1.1cm}
\end{abstract}

\maketitle

%\tableofcontents
%\comment{}

\section{\label{sec:Intro}Introduction}
Exploring the intricacies of condensed matter physics at the smallest scales has driven the pursuit of nanoscale sensors \cite{Spatial_Res, Rovny2024, Rembold2020,KangW}.
Spatial resolution is inherently dependent on the physical dimensions of the sensor unit. 
This makes sensors based on sfirst observed in a natural diamondingle defects exhibiting optically detected magnetic resonance (ODMR) \cite{first_ODMR}, commonly found in insulating materials like diamond and SiC, particularly advantageous\cite{Staudacher2013, ZaitsevsBook}.
At present, the field is dominated by diamond-based sensors utilizing nitrogen-vacancy (NV) centers, which have demonstrated exceptional performance across a range of sensing applications \cite{NVC1,NVC2,NVC3,NV_sense_mag,SNV_sense_el,SNV_sense_temp}. 
The development of NV centers embedded in fluorescence-collecting pillars on cantilevers has already progressed to a level, where they are commercially available \cite{Ren2024_Pillar, Qnami_Pillar}.
However, strong magnetic fields misaligned with the NV center's symmetry axis induce sub-level mixing, which disrupts spin initialization via spin pumping \cite{Tetienne}. 
This causes a loss of ODMR contrast, impairing the sensor’s functionality in the presence of strong magnetic fields with arbitrary orientations. For instance, a \SI{10}{\milli\tesla} field at an angle of \SI{10}{\degree}   causes a \SI{60}{\percent} reduction in sensitivity.
The TR12 center, an intrinsic defect in diamond with a sharp zero-phonon-line (ZPL) at \SI{471}{\nano\metre}, discovered in 1956 \cite{TR12_discovery}, was shown in a prior study to exhibit high ODMR contrast even in the presence of strong arbitrarily oriented magnetic fields \cite{Foglszinger2022}.
This makes it a viable substitute for NV centers in sensing scenarios involving strong magnetic fields.
Upon thorough investigation, it was found that the TR12 center lacks perfect photo-stability, which makes it less favourable for single-defect applications.
The TR12 center's ability to sense strong, arbitrarily oriented magnetic fields is primarily tied to its electronic level structure.
With this in mind, the ST2 center - short for Stuttgart - was proposed as a replacement. 
Discovered in 2019 in Stuttgart, it shares the same electronic level structure as the TR12 center, along with a sharp ZPL at \SI{446}{\nano\metre}, high ODMR contrast and a wide acceptance angle to strong magnetic fields. 
So far, no signs of photoinstability have been observed.

We conducted a comprehensive study on individual ST2 centers in diamond.
 A reliable method for creating ST2 centers is established.
The resulting centers were characterized in terms of their optical and magnetic properties, revealing their electronic level structure.
Additionally, we measured their sensing capabilities for magnetic and electric fields, as well as temperature, supported by a theoretical framework.
\section{Results}
\begin{figure*}
     \centering
     \includegraphics[width=\textwidth]{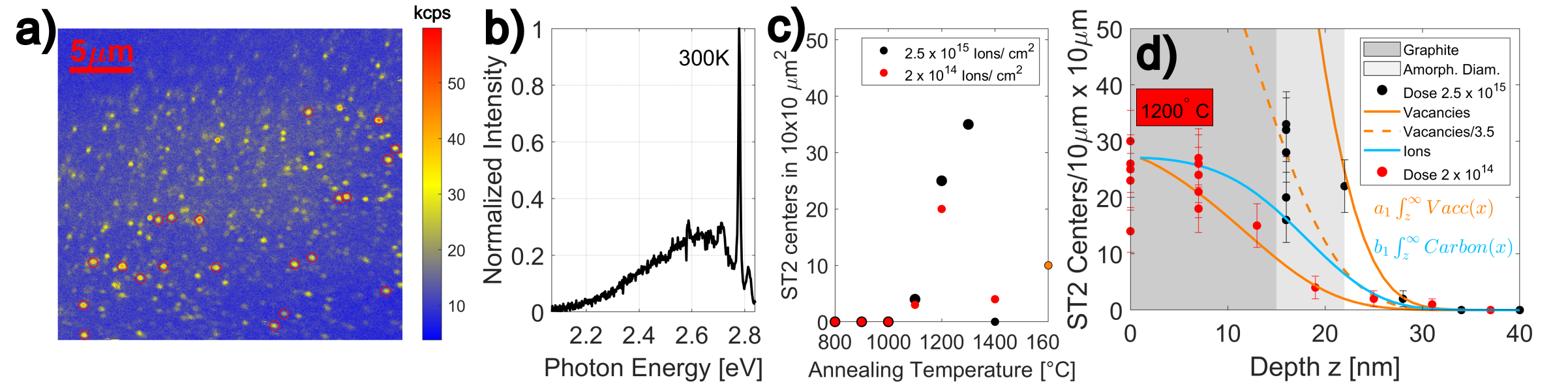}
        \caption{
     (a) Confocal scan image of the first artificial sample, covering a (\SI{27}{\micro\metre} $\times$ \SI{27}{\micro\metre}) region. Several ST2 centers have been identified.
        (b) Fluorescence spectrum of a single ST2 center, recorded at room temperature, within the natural diamond where the ST2 center was first discovered.
        (c) Temperature dependence of the ST2 creation yield for two different implantation doses, both at \SI{10}{keV} implantation energy. The spot at \SI{1600}{\celsius} is derived from implantation with multiple different energies, ranging from \SI{100}{keV} to \SI{500}{keV}.
        (d)  Depth profile of ST2 centers produced by one dose above and one below the graphitization threshold, followed by annealing at \SI{1200}{\celsius}. The solid orange lines, derived from SRIM-TRIM simulations using the same proportionality constant $a_1$, fit both observed ST2 center distributions. The cyan profile of implanted carbon ions, derived from the same SRIM-TRIM simulations and displayed only for the lower dose, does not fit the observed data.  At the dose above the graphitization threshold, an amorphous diamond layer was formed, inhibiting further ST2 center formation closer to the surface.
        }
        \label{fig:1,all}
\end{figure*}

\subsection{Creation of ST2 Centers}
ST2 centers were initially found in a natural diamond by observing their sharp zero-phonon line in emission and reaction of their fluorescence to a magnetic field.
The formation process of these centers was therefore unknown.
Later, a method of artificial creation of ST2 centers was revealed.
The identified protocol was to implant the diamond with $^{12}$C ions at levels slightly above the graphitization threshold, followed by annealing at \SI{1200}{\celsius}.
ST2 center creation was confirmed by locating individual centers in a confocal scan (Figure 1a) and comparing their fluorescence spectrum to that of centers from the natural diamond (Figure 1b).
After the initial finding, attention turned to improving the efficiency of ST2 center production due to the low yield observed in the first attempts.
In the following investigations, ST2 centers were never observed directly after implantation.
They only emerged after sample annealing at \SI{1100}{\celsius} or higher, as illustrated in Figure 1c.
When the annealing temperature exceeded \SI{1300}{\celsius}, slight graphitization of the diamond's surface made it challenging to evaluate the formation efficiency, due to the partial destruction of the center-containing layer.
Consequently, the observed reduction in yield at \SI{1400}{\celsius} versus \SI{1300}{\celsius} is derived from very limited data. At \SI{1600}{\celsius}, a single sample with higher implantation energies, confirmed a noticeable drop in overall efficiency compared to \SI{1200}{\celsius} and \SI{1300}{\celsius} despite a significant increase in overall implantation damage due to higher implantation energies. The highest yield was achieved after \SI{1300}{\celsius} annealing with around 35 Centers in a \SI{10}{\micro\metre}$\times$\SI{10}{\micro\metre} field.
To gain deeper insights into the formation process of ST2 centers, the sample was etched in incremental steps, evaluating the remaining concentration of ST2 centers at each stage.
The results, illustrated in Figure 1d, revealed that the distribution of ST2 centers corresponds closely with the total vacancy profile resulting from implantation as simulated using SRIM-TRIM \cite{Ziegler2010_TRIM}.
This suggests that the formation of ST2 centers is directly proportional to the implantation dose and independent of vacancy migration.
The exclusive use of carbon and the absence of  foreign elements for the implantation suggest that these centers are intrinsic defects, comprising only interstitial sites and vacancies within the diamond lattice.
This also fits the correlation between the concentration of ST2 centers and the vacancy profile.
\begin{figure*}
     \centering
	\includegraphics[width=\textwidth]{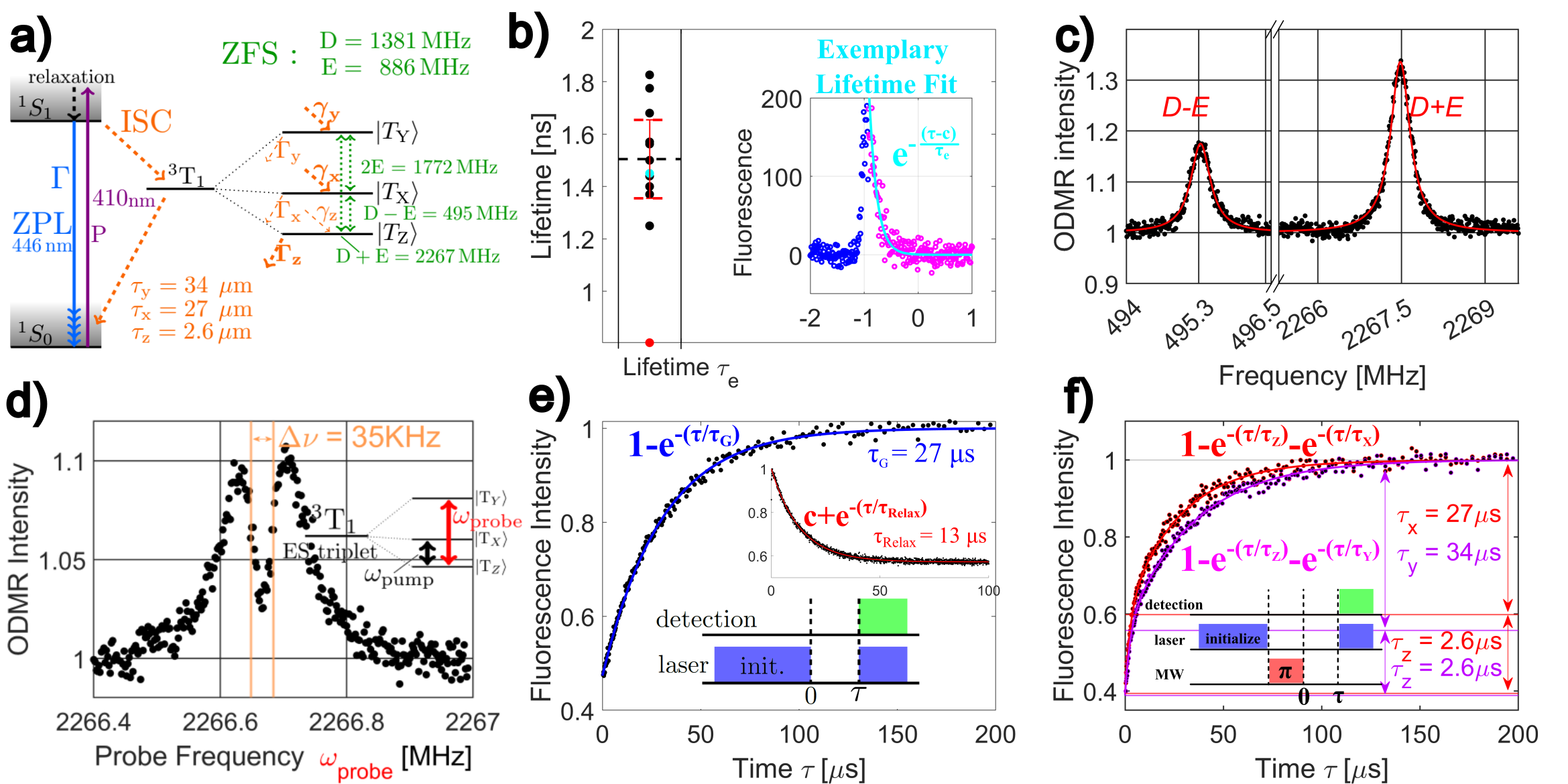}
             \label{fig:2,full}
             \caption{
             (a) Suggested level structure of the ST2 defect featuring a ground state singlet, an excited state singlet, and a metastable triplet state in between.
             (b) The excited singlet state lifetime was assessed for multiple centers (black dots). The mean value  (\SI{1.5}{\nano\second}, dashed black line) and the standard deviation (red errorbars) are shown, excluding the red-marked dataset. The inset shows the cyan-marked dataset, where the displayed dot corresponds to the fitted value of $\tau_e$.
             (c) CW ODMR spectroscopy in zero magnetic field (ZFS), displaying two resonances. A third resonance could not be observed.
             (d) Coherent population trapping measurement revealing the presence of two long-lived states, T$_x$ and T$_y$, with lifetimes of at least \SI{10}{\micro\second}. During this measurement one MW frequency $\omega_\text{pump}$ is fixed at the first ZFS resonance, while the other one, $\omega_\text{probe}$, is swept across the second ZFS resonance. Further details are available in the Supplementary Information \cite{Suppl}.
             (e) Decay of the total metastable population, revealing only a single lifetime $\tau_G$ (blue line) consistent with the long-lived states T$_x$ and T$_y$ (Figure 2f). The first inset illustrates the time-scale for fluorescence equilibration of the center after being initialized in the ground state by turning the laser off. The second inset displays the pulse sequence used to acquire the lifetime data.
             (f) Direct lifetime measurements for all triplet sublevels were performed by shifting the metastable population from one long-lived state into the short-lived T$_z$ state using a $\pi$-pulse derived from Rabi oscillations. The inset displays the pulse sequence used to acquire both datasets.
             }
\end{figure*}
Additionally, the lack of centers in deeper regions indicates that interstitials have to play a role, as vacancy migration might otherwise lead to these deeper centers.
However, a significant challenge arises from the relatively low proportionality constant between the implantation dose and the resulting ST2 center concentration of about \SI{2e-8}{centers/ion} for \SI{10}{keV} implantation energy. 
As the implantation dose increases, the sample starts to graphitize, forming amorphous diamond and non-diamond carbon phases where ST2 centers cannot form.
This theoretically limits the maximum achievable yield of ST2 centers with this method, effectively capping the local density of these defects at an estimated threshold of around \num{6e4} centers per cubic micron or 120 centers in a \SI{10}{\micro\metre}$\times$\SI{10}{\micro\metre} field for \SI{10}{keV} implantation \cite{Suppl}. 
A first attempt to optimize the yield lead to an increase from 30 to roughly 50 centers in 100 square microns, almost 50 percent of the theoretical maximum.
Given these relatively low numbers, the current production method presents challenges for the scalability and practical deployment of ST2 centers in high-density sensing scenarios. 
Even applications that rely on single defects are impacted by this issue, due to the inherent statistical nature of the production process and the extensive lattice damage formed around the defect.
\subsection{Optical Properties of ST2 Centers in Zero Magnetic Field }
\begin{figure*}
     \centering
         \includegraphics[width=\textwidth]{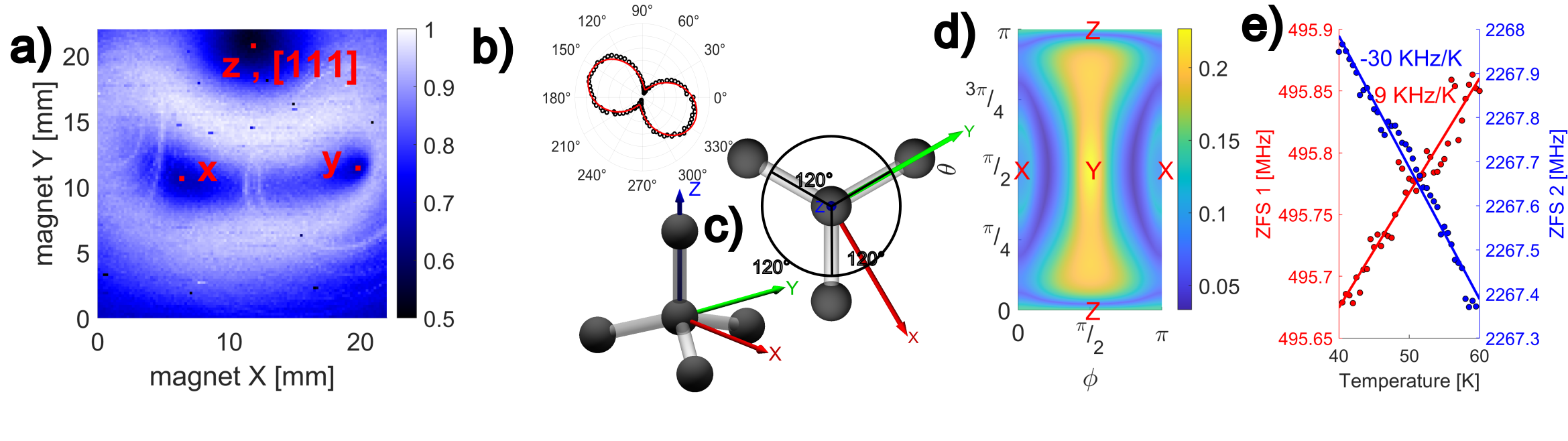}
     \caption{
     (a) Fluorescence response of a single ST2 center based on magnet position. Points where the magnetic field aligns with a sub-level of the metastable triplet state are emphasized. These points are found by comparison with simulations \cite{Suppl}. The faint rings arise due to cross relaxation mediated by nearby electron spins \cite{Suppl}.
     (b) Fluorescence dependence on the polarization of excitation light at low powers, revealing the characteristics of a dipole transition. Here one of the six dipole orientations is displayed.
     (c) Orientation of the principal axes of the metastable triplet state with respect to the diamond lattice, highlighting the threefold rotational symmetry inherent to the diamond lattice.
     (d) ODMR contrast simulation for ST2 as a function of magnetic field orientation at 30 mT.
         The figure shows the second highest ODMR contrast among the three available transitions which is the limiting factor for magnetic field sensitivity. The validity of this simulation is confirmed by the fact that it can reproduce the measured maps in Figure 3a. These simulation results provide the best overall illustrations of the sensing capacity of ST2 centers for stronger magnetic fields.
         (e) Temperature dependence of the ZFS from \SI{40}{\celsius} to \SI{60}{\celsius} for a single ST2 center. The shifts for the resonance correspond to shifts of \SI{10.5}{kHz/\kelvin} for D and \SI{-19.5}{kHz/\kelvin} for E. This suggests that ST2 centers can, in principle, also be used as temperature sensors. 
}
\end{figure*}
Due to the low yield of ST2 centers, they were observed as single defects exclusively.
Direct measurements of fluorescence decay under pulsed excitation revealed an excited state lifetime of approximately 1.5\,ns (Figure 2b).
The brightness of these defects, when optically saturated, was significantly affected by an external magnetic field, showing up to a twofold increase. 
The observed effect implies the presence of a sensitive spin state. 
Given that this effect occurs only during saturation, this spin state should be an excited state.
Based on earlier observations in molecules \cite{Joerg_early}, ST1 \cite{ST1} and TR12 \cite{Foglszinger2022}, the electronic level structure illustrated in Figure 2a was suggested as the simplest explanation for the results observed thus far. 
This model, which includes a single ground state, an excited singlet state, and a metastable triplet, has consistently matched all subsequent data.
We will therefore assume that this model is correct, while pointing out where our data serves as a consistency check.
It is crucial to highlight that the fluorescence observed originates only from the transition between the singlet states.
Consequently, the brightness of the centers is proportional to the total population in these singlets.
For different lifetimes of the triplet sub-levels, CW ODMR causes a shift of population from longer-lived to shorter-lived states, which shows up as up to three separate ODMR lines in the spectrum. Experimental data from multiple single defects confirm this, with two transitions being detected at \SI{495.3}{MHz} and \SI{2267.5}{MHz} (Figure 2c).
The ODMR contrast also disappears when the laser-power is reduced, indicating that the triplet is an excited state which is only significantly populated under saturation conditions.
The spin Hamiltonian of a triplet ($S=1$) state at zero magnetic field is represented by
$\text{H}=\mathrm{D}(S_z^2-S(S+1)/3)+\mathrm{E}(S_x^2-S_y^2)$.
The parameters $\mathrm{D}=\SI{1636.6}{MHz}$ and $\mathrm{E}=\SI{896.6}{MHz}$ can be deduced from the measured ODMR resonance frequencies as confirmed in the next subsection.
When microwave radiation is applied at both observed ODMR frequencies, the center displays coherent population trapping (CPT), as illustrated in Figure 2d.
This points to the existence of two long-lived states \cite{Suppl}.
It is clear from the width of the CPT resonance that these states have a lifetime of about \SI{10}{\micro\second} or longer.
The decay of the metastable population is directly observed by turning off the laser for a variable time $\tau$. 
As the laser excitation is restored, there is a resulting temporary increase in fluorescence as the population in the singlets is increased, as depicted in Figure 2e. 
The fit reveals only one decay time of roughly \SI{27}{\micro\second}.
It is reasonable to assume that the short-lived sub-level T$_z$ is minimally populated.
Consequently, the lifetimes of the two long-lived states can be observed in isolation by applying a microwave $\pi$-pulse, derived from Rabi oscillations, to shift the population from the long-lived states T$_x$ and T$_y$ to the short-lived T$_z$ (see Figure 2f).
The successful fitting of a double-exponential decay to this data, reveals all three lifetimes $\tau_x = \SI{27}{\micro\second}$, $\tau_y =\SI{34}{\micro\second}$ and $\tau_z = \SI{2.6}{\micro\second}$.
It confirms that the T$_z$ state is the least populated state when the center is excited continuously.
While the lifetime difference between T$_x$ and T$_y$ is minimal, it should lead to a weak ODMR signal when driving the transition between these states. 
Nonetheless, no ODMR signal has been observed at this frequency.
The most reasonable explanation is that the effect is canceled due to the disparity in population strength between these states.
Considering the transition rates into the metastable sub-levels T$_i$ ($\gamma_i$), and out of these sub-levels ($\Gamma_i$) as illustrated in Figure 2a, this requires the condition \(\gamma_x / \Gamma_x = \gamma_y / \Gamma_y\) to be fulfilled \cite{Suppl}.
However, directly assessing the rates into the metastable triplet states $\gamma_x$, $\gamma_y$ and $\gamma_z$  is not straightforward. 
It can only be assessed indirectly through simulations that reproduce the ODMR contrast of the resonance lines. This comparison clearly contradicts the above relation and instead leads to $\gamma_x=\gamma_y$, which is required to match the difference in ODMR contrast observed in Figure 2c.

The fluorescence intensity of the center at low excitation powers revealed the characteristics of a dipole transition when the linear polarization of the excitation laser is rotated (see Figure 3b).
Six distinct and recurring polarizations were observed for ST2 centers, one of which is shown in Figure 3b.
This finding offers an initial indication of the center's relatively low symmetry, which is further examined in the next subsection.
\subsection{\label{sec:Mag} Fluorescence Dependence on the Magnetic Field}
The previously noted dependence of the ST2 center fluorescence arises because the applied magnetic field induces mixing of the excited triplet state sub-levels. 
Consequently, their lifetimes become intertwined, impacting the overall dynamics and brightness of the center.
In this experiment, a permanent magnet (10x10x10\,mm$^3$ NdFeB, magnetization 1.4\,T perpendicular to the sample surface) is carefully positioned above the sample using stepper motors.
As the magnet's position changes, the fluorescence intensity of the center is recorded, covering various magnetic field orientations (see Figure 3a).
This data provides the means to evaluate the center's sensitivity to strong external magnetic fields and gives insights into its symmetry and orientation. 
The dark spots in Figure 3a result from the magnetic field being aligned with the principal axes of the triplet state, causing the corresponding sub-level to remain unaffected by the field.
By measuring ODMR spectra for these specific orientations, we can validate the applicability of the theory for Spin 1 systems as described by $\text{H}=\text{D}(\text{S}_z^2-S(S+1)/3) + \mathrm{E}(S_x^2-S_y^2) + g\SI{}{\micro_B} \textbf{S}\cdot\textbf{B} $ to ST2 centers, incorporating the zero-field splitting parameters D and E, the Bohr magneton $\SI{}{\micro_B}$, electron g-factor $g=2$ and $\mathrm{S}=1$ \cite{NV_sense_mag}. 
The ODMR frequencies for each magnetic field orientation can be derived from this Hamiltonian and its eigenstates.
By comparing these frequencies with the observed ODMR data, we can accurately assign the sub-level orientations to the dark spots in the magnetic maps in Figure 3a.
Matching this with NV data reveals that the center's z-orientation corresponds to the [111] direction in diamond.
Additionally, this allows for the assignment of zero-field transitions, which was assumed to be known in the preceding section.
The g-factor $g=2$ is validated by simulating the magnetic field strength at a particular distance from the magnet and comparing the findings to Figure 3a.
Additionally, simulating the rotation of the magnetic field based on the magnet's position helps to derive the exact x and y orientations in the laboratory frame.
As illustrated in Figure 3c, the $y$-axis of ST2 is situated in the plane formed by two adjacent $\sigma$-bonds.
Also, the entire system can be simulated to analyze the overall dynamics and reproduce the measured magnetic maps.
This simulation is primarily based on the level structure shown in Figure 2a and is described in greater detail in the supplementary material \cite{Suppl}.
The simulated maps align perfectly with the measured data. 
As this simulation incorporates all measured transition rates and the level structure, it offers a reliable confirmation of the model.
Based on these simulations, we also predict twelve species of differently oriented ST2 centers, organized into four triples \cite{Suppl}. 
These correspond to four distinct $\sigma$-bond orientations and the threefold rotational symmetry of the diamond lattice along each of these bonds.
Each orientation thereby gives its own characteristic magnetic map, most of which were confirmed in experiments \cite{Suppl}.  
The twelve maps align with the six different directions of the optical dipole, with each direction comprising two magnetic orientations.
The variety of orientations implies that the center must have a low symmetry group. 
With twelve different orientations, there are effectively only two symmetry candidates:
inversion symmetry C$_i$ and one mirror plane C$_s$. 
The lack of electric field sensitivity, which will be addressed later, indicates that the ST2 center exhibits inversion symmetry C$_i$.

\subsection{ST2 as Magnetic Field Sensors}
The results in Figure 3a, along with similar measurements, provide insights into the effectiveness of ST2 as a magnetic field sensor. 
However, the intertwined nature of magnetic field strength and orientation due to the permanent magnet makes it difficult to isolate these variables, complicating the interpretation of said data.
To make the situation more tangible, a simulation can be utilized in which magnetic field strength and orientation are separable.
The same model used to reproduce the measured maps in Figure 3a can also simulate the magnetic contrast of a single ST2 center. 
Further details on this simulation are provided in the supplementary information \cite{Suppl}.
The result of such a simulation is shown in Figure 3d. 
At a magnetic field strength of 30\,mT, a significant ODMR contrast is available across nearly the entire 4$\pi$ unit sphere. 
Given that the sensitivity is directly linked to the ODMR contrast of the center, this data confirms the suitability of ST2 centers as magnetic sensors, making them an excellent complement to NV centers.
They enable the detection of strong magnetic fields in almost any orientation with high ODMR contrast, while maintaining the typically high spatial resolution associated with single defects.
Exceptionally weak magnetic fields, where the induced frequency shift is comparable to or smaller than the zero-field splitting, are the only case where this does not apply.
In such scenarios, the high E value converts the transitions into clock transitions, as the ODMR frequency shows no linear dependence on the magnetic field.
This is particularly useful for temperature sensitivity and coherent population trapping (CPT) measurements, but it limits the magnetic sensitivity.
However, this is precisely the regime where the NV center excels, as the opposite logic holds true for NV.
Extremely weak magnetic fields do not cause significant state mixing, enabling the NV center to detect magnetic fields regardless of their orientation.
Thus, this regime is already accounted for by the NV.

Given the low yield of ST2 centers, only single-defect uses are viable.
The shot-noise limited magnetic sensitivity $\eta$ for such a single ST2 center can be approximated by considering the shot-noise $\sigma=10^2\,\mathrm{counts}/\sqrt{\mathrm{Hz}}$, the maximum gradient of the ODMR spectrum $C_{max}=\SI{1}{count/kHz}$, and the frequency shift of ODMR resonances $C_M=28\,\mathrm{GHz}/\mathrm{T}$:
\begin{equation}
\eta=\frac{\sigma}{C_{max}C_M}= 3.6 \frac{\SI{}{\micro T}}{\sqrt{\mathrm{Hz}}}
\end{equation}
The calculation is based on a conservative  5\,\% ODMR contrast superimposed on a average \SI{10}{kcps} fluorescence signal without any methods of fluorescence enhancement from a single ST2 center and a measured \SI{0.65}{MHz} linewidth. 
The maximum gradient $C_{max}$ equals the maximum derivative of a Lorentzian function with a full width at half maximum (FWHM) of \SI{0.65}{MHz} and an amplitude of $A=10^4 \cdot 0.05$ counts, which equals
\begin{equation}
C_{max}=\frac{A}{FWHM}\cdot \frac{3*\sqrt{3}}{4}=\SI{1}{count/kHz} .
\end{equation}
\subsection{ST2 as a General Sensing Platform}
To assess the broader capabilities of ST2 centers, we measured their sensitivity to electric fields and temperature.

To evaluate the electrical sensitivity of the ST2 center, we measured centers between two gold contacts connected to an external voltage source. 
However, no shift in ODMR frequencies was detected even at electric field strengths as high as $2*10^6$ V/m.
This observation aligns with the proposed inversion symmetry C$_i$ for ST2 centers, which would preclude the existence of a permanent electric dipole and thus first order electrical sensitivity.

Using the PID-controlled aluminum holder described in the methods section, ST2 ODMR frequencies were measured as a function of temperature between \SI{40}{\celsius} and \SI{60}{\celsius}. The temperature was stabilized for at least 5 minutes before each measurement.
As shown in Figure 3e, there is a clear linear dependence, with the two ODMR frequencies shifting by \SI{9}{kHz/\kelvin} and \SI{-30}{kHz/\kelvin}, respectively. 
This shift is two to three times lower than that observed for NV centers with a shift of $\SI{-74}{kHz/K}$ \cite{KangW}, suggesting that ST2 centers may not be preferable as primary temperature sensors in most applications. 
However, they in principal still show the potential to be used for local temperature assessment if convenient.

\section{\label{sec:CO} Conclusion and outlook}
ST2 centers can be produced in diamond using carbon implantation. 
However, the yield is quite low, with a conversion rate of approximately $2 \times 10^{-8}$ centers per implanted ion at \SI{10}{keV} implantation energy.
This rate results in a maximum achievable density of ST2 centers of approximately \num{6e4} centers per cubic micron.
Our findings reveal that the ST2 center features a sensitive spin state with coherent properties at room temperature. 
Through the combination of various measurements, we have outlined the full electronic level structure responsible for the dynamics of population within this spin state, as well as the overall properties of the center.
By examining the dependence of fluorescence on excitation polarization and external magnetic field, we identified twelve inequivalent orientations of ST2 centers. 
Since no electric sensitivity was observed, we conclude that the ST2 center exhibits $C_i$ symmetry.
ST2 complements the existing NV sensing capabilities exceptionally well, as its strength aligns precisely where the NV center has limitations, and vice versa. 
ST2 is particularly effective for measuring strong magnetic fields of any orientation with high spatial resolution and can also be used for temperature measurements under ambient conditions.
Figure 3d, which displays a notable ODMR contrast for arbitrary orientations of a \SI{30}{\milli\tesla} magnetic field, provides the best illustration of this point. In comparison, if the angle between the NV symmetry axis and a \SI{30}{\milli\tesla} magnetic field is greater than \SI{10}{\degree}, its ODMR contrast decreases below \SI{3}{\percent}.
Current efforts aim to transfer the NV center technology used in pillars on cantilevers to the ST2 center\cite{Ren2024_Pillar, Qnami_Pillar, Ali_Pillar}.
Preliminary results from these initial efforts are encouraging. 
The main challenge is the exceptionally low yield of ST2 centers using the current production method, which prevents the process from being economically scalable.
Addressing this issue is therefore the highest priority.
Furthermore, resolving this problem may facilitate progress toward achieving electroluminescence and electrical readout for ST2 centers.

\section{Methods}   
\subsection{Production}
ST2 centers (Figure 1a,b) were produced by implanting $^{12}$C ions at energies between \SI{5}{keV} and \SI{500}{keV} into the (100) plane of CVD diamond, followed by annealing at temperatures between \SI{800}{\celsius} and \SI{1600}{\celsius}.
The depth profile of these shallow-implanted centers was determined by incrementally etching the diamond surface using reactive ion etching (RIE) with $O_2$-plasma, while monitoring the concentration of remaining centers (Figure 1d).
Using optical lithography and electron beam physical vapor deposition (EBPVD), Au-contacts were deposited on the sample, facilitating the application of microwaves and DC voltages. The diodes were subsequently
connected to external measurement systems via wire bonding.
\subsection{Optical Characterization}
The defects in the prepared sample were analyzed using a home-built confocal microscope with 410 nm linearly polarized laser excitation (\cite{Suppl} Figure S1). 
Fluorescence emitted from the sample was directed to a single-photon detector, and a confocal scan was generated by moving the excitation point via a 3D piezo-stage. 
The excitation and detection points were kept coincident at all times.
In such confocal scans, the first step was to identify ST2 centers.
For this purpose, the fluorescence light was redirected to a spectrometer to record the optical spectrum, as shown in Figure 1b.
The sharp zero-phonon line (ZPL) at 446 nm served as the primary identification feature in these spectra.

The final step in verification involved measuring the ODMR spectrum, which confirmed the defect through the presence of two prominent resonance lines at \SI{495.3}{MHz} and \SI{2267.5}{MHz}.
\subsection{Characterization of Magnetic Properties}
To probe the magnetically susceptible spin states of the center, microwaves were applied to the sample through one of the integrated gold contacts.
Given that the spin states are excited states, as shown in the electronic level structure of ST2 (Figure 2a), it was critical to use a powerful laser to optically saturate the center for these measurements.
A permanent magnet was precisely moved into proximity of the sample using stepper motors, creating a magnetic field at the sample's position. 
This arrangement enabled the exploration of the centers' sensing capabilities in the presence of strong arbitrarily oriented magnetic fields.
\subsection{Thermal Sensitivity}
For a thorough evaluation of temperature sensitivity, the sample was mounted on a solid aluminum holder with an integrated PID-controlled heater.
The cool environment of the lab, combined with this set-up, ensured precise temperature control with a precision of less than \SI{0.1}{\kelvin}.
The aluminum holder thereby served as a thermal mass to stabilize the temperature.

\section*{Data availability}
The data that support the findings of this study are available from the corresponding author upon reasonable request, and will also be included in the final publication.

\section*{Code availability}

The MATLAB simulation code that supports the findings of this work is available from the corresponding author upon reasonable request.

\section*{Acknowledgments}
We thank Michelle Schweitzer for proofreading the manuscript. 
This work was supported by Bundesministerium für Bildung und Forschung (project UNIQ), Deutsche Forschungsgemeinschaft (grant KO4999/3-1), FET-Flagship Project SQUARE, EU project AMADEUS, DFG research group FOR 2724 and QTBW.
\section*{Author contributions}
The majority of the experimental work was conducted by J.F. at the University of Stuttgart under the supervision of R.K. and J.W.
Diamond samples were prepared mainly by J.F. and A.D. with contributions from G.V.A. and A.M.Z.
Data evaluation and simulations were performed by J.F., with contributions from all authors, especially R.K. .

\section*{COMPETING INTERESTS}
The authors declare no competing interests.

\nocite{*}
%%%%%%%%%%%%%%%%%%% BIBLIOGRAPHY
\bibliography{Literatur}% Produces the bibliography via BibTeX.

\begin{thebibliography}{30}%
\makeatletter
\providecommand \@ifxundefined [1]{%
 \@ifx{#1\undefined}
}%
\providecommand \@ifnum [1]{%
 \ifnum #1\expandafter \@firstoftwo
 \else \expandafter \@secondoftwo
 \fi
}%
\providecommand \@ifx [1]{%
 \ifx #1\expandafter \@firstoftwo
 \else \expandafter \@secondoftwo
 \fi
}%
\providecommand \natexlab [1]{#1}%
\providecommand \enquote  [1]{``#1''}%
\providecommand \bibnamefont  [1]{#1}%
\providecommand \bibfnamefont [1]{#1}%
\providecommand \citenamefont [1]{#1}%
\providecommand \href@noop [0]{\@secondoftwo}%
\providecommand \href [0]{\begingroup \@sanitize@url \@href}%
\providecommand \@href[1]{\@@startlink{#1}\@@href}%
\providecommand \@@href[1]{\endgroup#1\@@endlink}%
\providecommand \@sanitize@url [0]{\catcode `\\12\catcode `\$12\catcode
  `\&12\catcode `\#12\catcode `\^12\catcode `\_12\catcode `\%12\relax}%
\providecommand \@@startlink[1]{}%
\providecommand \@@endlink[0]{}%
\providecommand \url  [0]{\begingroup\@sanitize@url \@url }%
\providecommand \@url [1]{\endgroup\@href {#1}{\urlprefix }}%
\providecommand \urlprefix  [0]{URL }%
\providecommand \Eprint [0]{\href }%
\providecommand \doibase [0]{https://doi.org/}%
\providecommand \selectlanguage [0]{\@gobble}%
\providecommand \bibinfo  [0]{\@secondoftwo}%
\providecommand \bibfield  [0]{\@secondoftwo}%
\providecommand \translation [1]{[#1]}%
\providecommand \BibitemOpen [0]{}%
\providecommand \bibitemStop [0]{}%
\providecommand \bibitemNoStop [0]{.\EOS\space}%
\providecommand \EOS [0]{\spacefactor3000\relax}%
\providecommand \BibitemShut  [1]{\csname bibitem#1\endcsname}%
\let\auto@bib@innerbib\@empty
%</preamble>
\bibitem [{\citenamefont {Rembold}\ \emph
  {et~al.}(2020{\natexlab{a}})\citenamefont {Rembold}, \citenamefont {Oshnik},
  \citenamefont {Müller}, \citenamefont {Montangero}, \citenamefont
  {Calarco},\ and\ \citenamefont {Neu}}]{Spatial_Res}%
  \BibitemOpen
  \bibfield  {author} {\bibinfo {author} {\bibfnamefont {P.}~\bibnamefont
  {Rembold}}, \bibinfo {author} {\bibfnamefont {N.}~\bibnamefont {Oshnik}},
  \bibinfo {author} {\bibfnamefont {M.~M.}\ \bibnamefont {Müller}}, \bibinfo
  {author} {\bibfnamefont {S.}~\bibnamefont {Montangero}}, \bibinfo {author}
  {\bibfnamefont {T.}~\bibnamefont {Calarco}},\ and\ \bibinfo {author}
  {\bibfnamefont {E.}~\bibnamefont {Neu}},\ }\bibfield  {title} {\bibinfo
  {title} {Introduction to quantum optimal control for quantum sensing with
  nitrogen-vacancy centers in diamond},\ }\href
  {https://doi.org/10.1116/5.0006785} {\bibfield  {journal} {\bibinfo
  {journal} {AVS Quantum Science}\ }\textbf {\bibinfo {volume} {2}},\ \bibinfo
  {pages} {024701} (\bibinfo {year} {2020}{\natexlab{a}})}\BibitemShut
  {NoStop}%
\bibitem [{\citenamefont {Rovny}\ \emph {et~al.}(2024)\citenamefont {Rovny},
  \citenamefont {Gopalakrishnan}, \citenamefont {Jayich}, \citenamefont
  {Maletinsky}, \citenamefont {Demler},\ and\ \citenamefont
  {de~Leon}}]{Rovny2024}%
  \BibitemOpen
  \bibfield  {author} {\bibinfo {author} {\bibfnamefont {J.}~\bibnamefont
  {Rovny}}, \bibinfo {author} {\bibfnamefont {S.}~\bibnamefont
  {Gopalakrishnan}}, \bibinfo {author} {\bibfnamefont {A.~C.~B.}\ \bibnamefont
  {Jayich}}, \bibinfo {author} {\bibfnamefont {P.}~\bibnamefont {Maletinsky}},
  \bibinfo {author} {\bibfnamefont {E.}~\bibnamefont {Demler}},\ and\ \bibinfo
  {author} {\bibfnamefont {N.~P.}\ \bibnamefont {de~Leon}},\ }\bibfield
  {title} {\bibinfo {title} {New opportunities in condensed matter physics for
  nanoscale quantum sensors},\ }\href {http://arxiv.org/abs/2403.13710}
  {\bibfield  {journal} {\bibinfo  {journal} {arXiv [cond-mat.mes-hall]}\ }
  (\bibinfo {year} {2024})},\ \Eprint {https://arxiv.org/abs/2403.13710}
  {arXiv:2403.13710} \BibitemShut {NoStop}%
\bibitem [{\citenamefont {Rembold}\ \emph
  {et~al.}(2020{\natexlab{b}})\citenamefont {Rembold}, \citenamefont {Oshnik},
  \citenamefont {M{\"u}ller}, \citenamefont {Montangero}, \citenamefont
  {Calarco},\ and\ \citenamefont {Neu}}]{Rembold2020}%
  \BibitemOpen
  \bibfield  {author} {\bibinfo {author} {\bibfnamefont {P.}~\bibnamefont
  {Rembold}}, \bibinfo {author} {\bibfnamefont {N.}~\bibnamefont {Oshnik}},
  \bibinfo {author} {\bibfnamefont {M.~M.}\ \bibnamefont {M{\"u}ller}},
  \bibinfo {author} {\bibfnamefont {S.}~\bibnamefont {Montangero}}, \bibinfo
  {author} {\bibfnamefont {T.}~\bibnamefont {Calarco}},\ and\ \bibinfo {author}
  {\bibfnamefont {E.}~\bibnamefont {Neu}},\ }\bibfield  {title} {\bibinfo
  {title} {Introduction to quantum optimal control for quantum sensing with
  nitrogen-vacancy centers in diamond},\ }\href
  {https://doi.org/10.1116/5.0006785} {\bibfield  {journal} {\bibinfo
  {journal} {AVS Quantum Science}\ }\textbf {\bibinfo {volume} {2}},\ \bibinfo
  {pages} {024701} (\bibinfo {year} {2020}{\natexlab{b}})}\BibitemShut
  {NoStop}%
\bibitem [{\citenamefont {Liu}\ \emph {et~al.}(2021)\citenamefont {Liu},
  \citenamefont {Leong}, \citenamefont {Xia}, \citenamefont {Feng},
  \citenamefont {Finkler}, \citenamefont {Denisenko}, \citenamefont
  {Wrachtrup}, \citenamefont {Li},\ and\ \citenamefont {Liu}}]{KangW}%
  \BibitemOpen
  \bibfield  {author} {\bibinfo {author} {\bibfnamefont {C.-F.}\ \bibnamefont
  {Liu}}, \bibinfo {author} {\bibfnamefont {W.-H.}\ \bibnamefont {Leong}},
  \bibinfo {author} {\bibfnamefont {K.}~\bibnamefont {Xia}}, \bibinfo {author}
  {\bibfnamefont {X.}~\bibnamefont {Feng}}, \bibinfo {author} {\bibfnamefont
  {A.}~\bibnamefont {Finkler}}, \bibinfo {author} {\bibfnamefont
  {A.}~\bibnamefont {Denisenko}}, \bibinfo {author} {\bibfnamefont
  {J.}~\bibnamefont {Wrachtrup}}, \bibinfo {author} {\bibfnamefont
  {Q.}~\bibnamefont {Li}},\ and\ \bibinfo {author} {\bibfnamefont {R.-B.}\
  \bibnamefont {Liu}},\ }\bibfield  {title} {\bibinfo {title} {Ultra-sensitive
  hybrid diamond nanothermometer},\ }\href
  {https://doi.org/10.1093/nsr/nwaa194} {\bibfield  {journal} {\bibinfo
  {journal} {National Science Review}\ }\textbf {\bibinfo {volume} {8}},\
  \bibinfo {pages} {nwaa194} (\bibinfo {year} {2021})}\BibitemShut {NoStop}%
\bibitem [{\citenamefont {Brossel}\ and\ \citenamefont
  {Bitter}(1952)}]{first_ODMR}%
  \BibitemOpen
  \bibfield  {author} {\bibinfo {author} {\bibfnamefont {J.}~\bibnamefont
  {Brossel}}\ and\ \bibinfo {author} {\bibfnamefont {F.}~\bibnamefont
  {Bitter}},\ }\bibfield  {title} {\bibinfo {title} {A new "double resonance"
  method for investigating atomic energy levels. application to $\mathrm{Hg}
  ^{3}p_{1}$},\ }\href {https://doi.org/10.1103/PhysRev.86.308} {\bibfield
  {journal} {\bibinfo  {journal} {Phys. Rev.}\ }\textbf {\bibinfo {volume}
  {86}},\ \bibinfo {pages} {308} (\bibinfo {year} {1952})}\BibitemShut
  {NoStop}%
\bibitem [{\citenamefont {Staudacher}\ \emph {et~al.}(2013)\citenamefont
  {Staudacher}, \citenamefont {Shi}, \citenamefont {Pezzagna}, \citenamefont
  {Meijer}, \citenamefont {Du}, \citenamefont {Meriles}, \citenamefont
  {Reinhard},\ and\ \citenamefont {Wrachtrup}}]{Staudacher2013}%
  \BibitemOpen
  \bibfield  {author} {\bibinfo {author} {\bibfnamefont {T.}~\bibnamefont
  {Staudacher}}, \bibinfo {author} {\bibfnamefont {F.}~\bibnamefont {Shi}},
  \bibinfo {author} {\bibfnamefont {S.}~\bibnamefont {Pezzagna}}, \bibinfo
  {author} {\bibfnamefont {J.}~\bibnamefont {Meijer}}, \bibinfo {author}
  {\bibfnamefont {J.}~\bibnamefont {Du}}, \bibinfo {author} {\bibfnamefont
  {C.~A.}\ \bibnamefont {Meriles}}, \bibinfo {author} {\bibfnamefont
  {F.}~\bibnamefont {Reinhard}},\ and\ \bibinfo {author} {\bibfnamefont
  {J.}~\bibnamefont {Wrachtrup}},\ }\bibfield  {title} {\bibinfo {title}
  {Nuclear magnetic resonance spectroscopy on a (5-nanometer)3 sample volume},\
  }\href {https://doi.org/10.1126/science.1231675} {\bibfield  {journal}
  {\bibinfo  {journal} {Science}\ }\textbf {\bibinfo {volume} {339}},\ \bibinfo
  {pages} {561} (\bibinfo {year} {2013})}\BibitemShut {NoStop}%
\bibitem [{\citenamefont {Zaitsev}(2001)}]{ZaitsevsBook}%
  \BibitemOpen
  \bibfield  {author} {\bibinfo {author} {\bibfnamefont {A.}~\bibnamefont
  {Zaitsev}},\ }\href@noop {} {\emph {\bibinfo {title} {Optical Properties of
  Diamond}}},\ \bibinfo {edition} {1st}\ ed.\ (\bibinfo  {publisher}
  {Springer},\ \bibinfo {year} {2001})\BibitemShut {NoStop}%
\bibitem [{\citenamefont {Acosta}\ \emph {et~al.}(2012)\citenamefont {Acosta},
  \citenamefont {Santori}, \citenamefont {Faraon}, \citenamefont {Huang},
  \citenamefont {Fu}, \citenamefont {Stacey}, \citenamefont {Simpson},
  \citenamefont {Ganesan}, \citenamefont {Tomljenovic-Hanic}, \citenamefont
  {Greentree}, \citenamefont {Prawer},\ and\ \citenamefont
  {Beausoleil}}]{NVC1}%
  \BibitemOpen
  \bibfield  {author} {\bibinfo {author} {\bibfnamefont {V.~M.}\ \bibnamefont
  {Acosta}}, \bibinfo {author} {\bibfnamefont {C.}~\bibnamefont {Santori}},
  \bibinfo {author} {\bibfnamefont {A.}~\bibnamefont {Faraon}}, \bibinfo
  {author} {\bibfnamefont {Z.}~\bibnamefont {Huang}}, \bibinfo {author}
  {\bibfnamefont {K.-M.~C.}\ \bibnamefont {Fu}}, \bibinfo {author}
  {\bibfnamefont {A.}~\bibnamefont {Stacey}}, \bibinfo {author} {\bibfnamefont
  {D.~A.}\ \bibnamefont {Simpson}}, \bibinfo {author} {\bibfnamefont
  {K.}~\bibnamefont {Ganesan}}, \bibinfo {author} {\bibfnamefont
  {S.}~\bibnamefont {Tomljenovic-Hanic}}, \bibinfo {author} {\bibfnamefont
  {A.~D.}\ \bibnamefont {Greentree}}, \bibinfo {author} {\bibfnamefont
  {S.}~\bibnamefont {Prawer}},\ and\ \bibinfo {author} {\bibfnamefont {R.~G.}\
  \bibnamefont {Beausoleil}},\ }\bibfield  {title} {\bibinfo {title} {Dynamic
  stabilization of the optical resonances of single nitrogen-vacancy centers in
  diamond},\ }\href {https://doi.org/10.1103/PhysRevLett.108.206401} {\bibfield
   {journal} {\bibinfo  {journal} {Phys. Rev. Lett.}\ }\textbf {\bibinfo
  {volume} {108}},\ \bibinfo {pages} {206401} (\bibinfo {year}
  {2012})}\BibitemShut {NoStop}%
\bibitem [{\citenamefont {Fuchs}\ \emph {et~al.}(2010)\citenamefont {Fuchs},
  \citenamefont {Dobrovitski}, \citenamefont {Toyli}, \citenamefont {Heremans},
  \citenamefont {Weis}, \citenamefont {Schenkel},\ and\ \citenamefont
  {Awschalom}}]{NVC2}%
  \BibitemOpen
  \bibfield  {author} {\bibinfo {author} {\bibfnamefont {G.~D.}\ \bibnamefont
  {Fuchs}}, \bibinfo {author} {\bibfnamefont {V.~V.}\ \bibnamefont
  {Dobrovitski}}, \bibinfo {author} {\bibfnamefont {D.~M.}\ \bibnamefont
  {Toyli}}, \bibinfo {author} {\bibfnamefont {F.~J.}\ \bibnamefont {Heremans}},
  \bibinfo {author} {\bibfnamefont {C.~D.}\ \bibnamefont {Weis}}, \bibinfo
  {author} {\bibfnamefont {T.}~\bibnamefont {Schenkel}},\ and\ \bibinfo
  {author} {\bibfnamefont {D.~D.}\ \bibnamefont {Awschalom}},\ }\bibfield
  {title} {\bibinfo {title} {Excited-state spin coherence of a single
  nitrogen--vacancy centre in diamond},\ }\href
  {https://doi.org/10.1038/nphys1716} {\bibfield  {journal} {\bibinfo
  {journal} {Nature Physics}\ }\textbf {\bibinfo {volume} {6}},\ \bibinfo
  {pages} {668} (\bibinfo {year} {2010})}\BibitemShut {NoStop}%
\bibitem [{\citenamefont {Shen}\ \emph {et~al.}(2008)\citenamefont {Shen},
  \citenamefont {Sweeney},\ and\ \citenamefont {Wang}}]{NVC3}%
  \BibitemOpen
  \bibfield  {author} {\bibinfo {author} {\bibfnamefont {Y.}~\bibnamefont
  {Shen}}, \bibinfo {author} {\bibfnamefont {T.~M.}\ \bibnamefont {Sweeney}},\
  and\ \bibinfo {author} {\bibfnamefont {H.}~\bibnamefont {Wang}},\ }\bibfield
  {title} {\bibinfo {title} {Zero-phonon linewidth of single nitrogen vacancy
  centers in diamond nanocrystals},\ }\href
  {https://doi.org/10.1103/PhysRevB.77.033201} {\bibfield  {journal} {\bibinfo
  {journal} {Phys. Rev. B}\ }\textbf {\bibinfo {volume} {77}},\ \bibinfo
  {pages} {033201} (\bibinfo {year} {2008})}\BibitemShut {NoStop}%
\bibitem [{\citenamefont {Balasubramanian}\ \emph {et~al.}(2008)\citenamefont
  {Balasubramanian}, \citenamefont {Chan}, \citenamefont {Kolesov},
  \citenamefont {Al-Hmoud}, \citenamefont {Tisler}, \citenamefont {Shin},
  \citenamefont {Kim}, \citenamefont {Wojcik}, \citenamefont {Hemmer},
  \citenamefont {Krueger}, \citenamefont {Hanke}, \citenamefont
  {Leitenstorfer}, \citenamefont {Bratschitsch}, \citenamefont {Jelezko},\ and\
  \citenamefont {Wrachtrup}}]{NV_sense_mag}%
  \BibitemOpen
  \bibfield  {author} {\bibinfo {author} {\bibfnamefont {G.}~\bibnamefont
  {Balasubramanian}}, \bibinfo {author} {\bibfnamefont {I.~Y.}\ \bibnamefont
  {Chan}}, \bibinfo {author} {\bibfnamefont {R.}~\bibnamefont {Kolesov}},
  \bibinfo {author} {\bibfnamefont {M.}~\bibnamefont {Al-Hmoud}}, \bibinfo
  {author} {\bibfnamefont {J.}~\bibnamefont {Tisler}}, \bibinfo {author}
  {\bibfnamefont {C.}~\bibnamefont {Shin}}, \bibinfo {author} {\bibfnamefont
  {C.}~\bibnamefont {Kim}}, \bibinfo {author} {\bibfnamefont {A.}~\bibnamefont
  {Wojcik}}, \bibinfo {author} {\bibfnamefont {P.~R.}\ \bibnamefont {Hemmer}},
  \bibinfo {author} {\bibfnamefont {A.}~\bibnamefont {Krueger}}, \bibinfo
  {author} {\bibfnamefont {T.}~\bibnamefont {Hanke}}, \bibinfo {author}
  {\bibfnamefont {A.}~\bibnamefont {Leitenstorfer}}, \bibinfo {author}
  {\bibfnamefont {R.}~\bibnamefont {Bratschitsch}}, \bibinfo {author}
  {\bibfnamefont {F.}~\bibnamefont {Jelezko}},\ and\ \bibinfo {author}
  {\bibfnamefont {J.}~\bibnamefont {Wrachtrup}},\ }\bibfield  {title} {\bibinfo
  {title} {Nanoscale imaging magnetometry with diamond spins under ambient
  conditions},\ }\href {https://doi.org/10.1038/nature07278} {\bibfield
  {journal} {\bibinfo  {journal} {Nature}\ }\textbf {\bibinfo {volume} {455}},\
  \bibinfo {pages} {648} (\bibinfo {year} {2008})}\BibitemShut {NoStop}%
\bibitem [{\citenamefont {Dolde}\ \emph {et~al.}(2011)\citenamefont {Dolde},
  \citenamefont {Fedder}, \citenamefont {Doherty}, \citenamefont {N{\"o}bauer},
  \citenamefont {Rempp}, \citenamefont {Balasubramanian}, \citenamefont {Wolf},
  \citenamefont {Reinhard}, \citenamefont {Hollenberg}, \citenamefont
  {Jelezko},\ and\ \citenamefont {Wrachtrup}}]{SNV_sense_el}%
  \BibitemOpen
  \bibfield  {author} {\bibinfo {author} {\bibfnamefont {F.}~\bibnamefont
  {Dolde}}, \bibinfo {author} {\bibfnamefont {H.}~\bibnamefont {Fedder}},
  \bibinfo {author} {\bibfnamefont {M.~W.}\ \bibnamefont {Doherty}}, \bibinfo
  {author} {\bibfnamefont {T.}~\bibnamefont {N{\"o}bauer}}, \bibinfo {author}
  {\bibfnamefont {F.}~\bibnamefont {Rempp}}, \bibinfo {author} {\bibfnamefont
  {G.}~\bibnamefont {Balasubramanian}}, \bibinfo {author} {\bibfnamefont
  {T.}~\bibnamefont {Wolf}}, \bibinfo {author} {\bibfnamefont {F.}~\bibnamefont
  {Reinhard}}, \bibinfo {author} {\bibfnamefont {L.~C.~L.}\ \bibnamefont
  {Hollenberg}}, \bibinfo {author} {\bibfnamefont {F.}~\bibnamefont
  {Jelezko}},\ and\ \bibinfo {author} {\bibfnamefont {J.}~\bibnamefont
  {Wrachtrup}},\ }\bibfield  {title} {\bibinfo {title} {Electric-field sensing
  using single diamond spins},\ }\href {https://doi.org/10.1038/nphys1969}
  {\bibfield  {journal} {\bibinfo  {journal} {Nature Physics}\ }\textbf
  {\bibinfo {volume} {7}},\ \bibinfo {pages} {459} (\bibinfo {year}
  {2011})}\BibitemShut {NoStop}%
\bibitem [{\citenamefont {Kucsko}\ \emph {et~al.}(2013)\citenamefont {Kucsko},
  \citenamefont {Maurer}, \citenamefont {Yao}, \citenamefont {Kubo},
  \citenamefont {Noh}, \citenamefont {Lo}, \citenamefont {Park},\ and\
  \citenamefont {Lukin}}]{SNV_sense_temp}%
  \BibitemOpen
  \bibfield  {author} {\bibinfo {author} {\bibfnamefont {G.}~\bibnamefont
  {Kucsko}}, \bibinfo {author} {\bibfnamefont {P.~C.}\ \bibnamefont {Maurer}},
  \bibinfo {author} {\bibfnamefont {N.~Y.}\ \bibnamefont {Yao}}, \bibinfo
  {author} {\bibfnamefont {M.}~\bibnamefont {Kubo}}, \bibinfo {author}
  {\bibfnamefont {H.~J.}\ \bibnamefont {Noh}}, \bibinfo {author} {\bibfnamefont
  {P.~K.}\ \bibnamefont {Lo}}, \bibinfo {author} {\bibfnamefont
  {H.}~\bibnamefont {Park}},\ and\ \bibinfo {author} {\bibfnamefont {M.~D.}\
  \bibnamefont {Lukin}},\ }\bibfield  {title} {\bibinfo {title}
  {Nanometre-scale thermometry in a living cell},\ }\href
  {https://doi.org/10.1038/nature12373} {\bibfield  {journal} {\bibinfo
  {journal} {Nature}\ }\textbf {\bibinfo {volume} {500}},\ \bibinfo {pages}
  {54} (\bibinfo {year} {2013})}\BibitemShut {NoStop}%
\bibitem [{\citenamefont {Ren}\ and\ \citenamefont
  {Takahashi}(2024)}]{Ren2024_Pillar}%
  \BibitemOpen
  \bibfield  {author} {\bibinfo {author} {\bibfnamefont {Y.}~\bibnamefont
  {Ren}}\ and\ \bibinfo {author} {\bibfnamefont {S.}~\bibnamefont
  {Takahashi}},\ }\bibfield  {title} {\bibinfo {title} {Detection of electron
  paramagnetic resonance of two electron spins using a single nv center in
  diamond},\ }\href {https://doi.org/10.1063/5.0224013} {\bibfield  {journal}
  {\bibinfo  {journal} {APL Quantum}\ }\textbf {\bibinfo {volume} {1}},\
  \bibinfo {pages} {046108} (\bibinfo {year} {2024})}\BibitemShut {NoStop}%
\bibitem [{\citenamefont {{Qnami}}()}]{Qnami_Pillar}%
  \BibitemOpen
  \bibfield  {author} {\bibinfo {author} {\bibnamefont {{Qnami}}},\ }\href@noop
  {} {\bibinfo {title} {Quantum sensing leaders in nanoscale precision}},\
  \bibinfo {howpublished}
  {\url{https://qnami.ch/wp-content/uploads/2021/03/Quantum-Foundry-Case-Study-16.03.2021.pdf}}\BibitemShut
  {NoStop}%
\bibitem [{\citenamefont {Tetienne1}\ \emph {et~al.}(2012)\citenamefont
  {Tetienne1}, \citenamefont {Rondin}, \citenamefont {Spinicelli},
  \citenamefont {Chipaux}, \citenamefont {Debuisschert}, \citenamefont {Roch},\
  and\ \citenamefont {Jacques}}]{Tetienne}%
  \BibitemOpen
  \bibfield  {author} {\bibinfo {author} {\bibfnamefont {J.-P.}\ \bibnamefont
  {Tetienne1}}, \bibinfo {author} {\bibfnamefont {L.}~\bibnamefont {Rondin}},
  \bibinfo {author} {\bibfnamefont {P.}~\bibnamefont {Spinicelli}}, \bibinfo
  {author} {\bibfnamefont {M.}~\bibnamefont {Chipaux}}, \bibinfo {author}
  {\bibfnamefont {T.}~\bibnamefont {Debuisschert}}, \bibinfo {author}
  {\bibfnamefont {J.-F.}\ \bibnamefont {Roch}},\ and\ \bibinfo {author}
  {\bibfnamefont {V.}~\bibnamefont {Jacques}},\ }\bibfield  {title} {\bibinfo
  {title} {Magnetic-field-dependent photodynamics of single nv defects in
  diamond: an application to qualitative all-optical magnetic imaging},\
  }\href@noop {} {\bibfield  {journal} {\bibinfo  {journal} {New Journal of
  Physics}\ }\textbf {\bibinfo {volume} {14}} (\bibinfo {year}
  {2012})}\BibitemShut {NoStop}%
\bibitem [{\citenamefont {C.D.}\ \emph {et~al.}(1956)\citenamefont {C.D.},
  \citenamefont {Ditchburn},\ and\ \citenamefont {Dyer}}]{TR12_discovery}%
  \BibitemOpen
  \bibfield  {author} {\bibinfo {author} {\bibfnamefont {C.}~\bibnamefont
  {C.D.}}, \bibinfo {author} {\bibfnamefont {R.}~\bibnamefont {Ditchburn}},\
  and\ \bibinfo {author} {\bibfnamefont {H.}~\bibnamefont {Dyer}},\ }\bibfield
  {title} {\bibinfo {title} {“the absorption spectra of natural and
  irradiated diamonds.”},\ }\href {www.jstor.org/stable/99841} {\bibfield
  {journal} {\bibinfo  {journal} {Proceedings of the Royal Society of London.
  Series A, Mathematical and Physical Sciences}\ }\textbf {\bibinfo {volume}
  {234}},\ \bibinfo {pages} {363} (\bibinfo {year} {1956})}\BibitemShut
  {NoStop}%
\bibitem [{\citenamefont {Foglszinger}\ \emph {et~al.}(2022)\citenamefont
  {Foglszinger}, \citenamefont {Denisenko}, \citenamefont {Kornher},
  \citenamefont {Schreck}, \citenamefont {Knolle}, \citenamefont {Yavkin},
  \citenamefont {Kolesov},\ and\ \citenamefont {Wrachtrup}}]{Foglszinger2022}%
  \BibitemOpen
  \bibfield  {author} {\bibinfo {author} {\bibfnamefont {J.}~\bibnamefont
  {Foglszinger}}, \bibinfo {author} {\bibfnamefont {A.}~\bibnamefont
  {Denisenko}}, \bibinfo {author} {\bibfnamefont {T.}~\bibnamefont {Kornher}},
  \bibinfo {author} {\bibfnamefont {M.}~\bibnamefont {Schreck}}, \bibinfo
  {author} {\bibfnamefont {W.}~\bibnamefont {Knolle}}, \bibinfo {author}
  {\bibfnamefont {B.}~\bibnamefont {Yavkin}}, \bibinfo {author} {\bibfnamefont
  {R.}~\bibnamefont {Kolesov}},\ and\ \bibinfo {author} {\bibfnamefont
  {J.}~\bibnamefont {Wrachtrup}},\ }\bibfield  {title} {\bibinfo {title} {Tr12
  centers in diamond as a room temperature atomic scale vector magnetometer},\
  }\href {https://doi.org/10.1038/s41534-022-00566-8} {\bibfield  {journal}
  {\bibinfo  {journal} {npj Quantum Information}\ }\textbf {\bibinfo {volume}
  {8}},\ \bibinfo {pages} {65} (\bibinfo {year} {2022})}\BibitemShut {NoStop}%
\bibitem [{\citenamefont {Ziegler}\ \emph {et~al.}(2010)\citenamefont
  {Ziegler}, \citenamefont {Ziegler},\ and\ \citenamefont
  {Biersack}}]{Ziegler2010_TRIM}%
  \BibitemOpen
  \bibfield  {author} {\bibinfo {author} {\bibfnamefont {J.~F.}\ \bibnamefont
  {Ziegler}}, \bibinfo {author} {\bibfnamefont {M.~D.}\ \bibnamefont
  {Ziegler}},\ and\ \bibinfo {author} {\bibfnamefont {J.~P.}\ \bibnamefont
  {Biersack}},\ }\bibfield  {title} {\bibinfo {title} {Srim -- the stopping and
  range of ions in matter (2010)},\ }\href
  {https://www.sciencedirect.com/science/article/pii/S0168583X10001862}
  {\bibfield  {journal} {\bibinfo  {journal} {Nuclear Instruments and Methods
  in Physics Research Section B: Beam Interactions with Materials and Atoms}\
  }\textbf {\bibinfo {volume} {268}},\ \bibinfo {pages} {1818} (\bibinfo {year}
  {2010})}\BibitemShut {NoStop}%
\bibitem [{\citenamefont {{Me-Myself and I}}()}]{Suppl}%
  \BibitemOpen
  \bibfield  {author} {\bibinfo {author} {\bibnamefont {{Me-Myself and I}}},\
  }\href@noop {} {\bibinfo {title} {Placeholder for supplementary
  info}}\BibitemShut {NoStop}%
\bibitem [{\citenamefont {Wrachtrup}\ \emph {et~al.}(1993)\citenamefont
  {Wrachtrup}, \citenamefont {von Borczyskowski}, \citenamefont {Bernard},
  \citenamefont {Orrit},\ and\ \citenamefont {Brown}}]{Joerg_early}%
  \BibitemOpen
  \bibfield  {author} {\bibinfo {author} {\bibfnamefont {J.}~\bibnamefont
  {Wrachtrup}}, \bibinfo {author} {\bibfnamefont {C.}~\bibnamefont {von
  Borczyskowski}}, \bibinfo {author} {\bibfnamefont {J.}~\bibnamefont
  {Bernard}}, \bibinfo {author} {\bibfnamefont {M.}~\bibnamefont {Orrit}},\
  and\ \bibinfo {author} {\bibfnamefont {R.}~\bibnamefont {Brown}},\ }\bibfield
   {title} {\bibinfo {title} {Optically detected spin coherence of single
  molecules},\ }\href {https://doi.org/10.1103/PhysRevLett.71.3565} {\bibfield
  {journal} {\bibinfo  {journal} {Phys. Rev. Lett.}\ }\textbf {\bibinfo
  {volume} {71}},\ \bibinfo {pages} {3565} (\bibinfo {year}
  {1993})}\BibitemShut {NoStop}%
\bibitem [{\citenamefont {Lee}\ \emph {et~al.}(2013)\citenamefont {Lee},
  \citenamefont {Widmann}, \citenamefont {Rendler}, \citenamefont {Doherty},
  \citenamefont {Babinec}, \citenamefont {Yang}, \citenamefont {Eyer},
  \citenamefont {Siyushev}, \citenamefont {Hausmann}, \citenamefont {Loncar},
  \citenamefont {Bodrog}, \citenamefont {Gali}, \citenamefont {Manson},
  \citenamefont {Fedder},\ and\ \citenamefont {Wrachtrup}}]{ST1}%
  \BibitemOpen
  \bibfield  {author} {\bibinfo {author} {\bibfnamefont {S.-Y.}\ \bibnamefont
  {Lee}}, \bibinfo {author} {\bibfnamefont {M.}~\bibnamefont {Widmann}},
  \bibinfo {author} {\bibfnamefont {T.}~\bibnamefont {Rendler}}, \bibinfo
  {author} {\bibfnamefont {M.~W.}\ \bibnamefont {Doherty}}, \bibinfo {author}
  {\bibfnamefont {T.~M.}\ \bibnamefont {Babinec}}, \bibinfo {author}
  {\bibfnamefont {S.}~\bibnamefont {Yang}}, \bibinfo {author} {\bibfnamefont
  {M.}~\bibnamefont {Eyer}}, \bibinfo {author} {\bibfnamefont {P.}~\bibnamefont
  {Siyushev}}, \bibinfo {author} {\bibfnamefont {B.~J.~M.}\ \bibnamefont
  {Hausmann}}, \bibinfo {author} {\bibfnamefont {M.}~\bibnamefont {Loncar}},
  \bibinfo {author} {\bibfnamefont {Z.}~\bibnamefont {Bodrog}}, \bibinfo
  {author} {\bibfnamefont {A.}~\bibnamefont {Gali}}, \bibinfo {author}
  {\bibfnamefont {N.~B.}\ \bibnamefont {Manson}}, \bibinfo {author}
  {\bibfnamefont {H.}~\bibnamefont {Fedder}},\ and\ \bibinfo {author}
  {\bibfnamefont {J.}~\bibnamefont {Wrachtrup}},\ }\bibfield  {title} {\bibinfo
  {title} {Readout and control of a single nuclear spin with a metastable
  electron spin ancilla},\ }\href {https://doi.org/10.1038/nnano.2013.104}
  {\bibfield  {journal} {\bibinfo  {journal} {Nature Nanotechnology}\ }\textbf
  {\bibinfo {volume} {8}},\ \bibinfo {pages} {487} (\bibinfo {year}
  {2013})}\BibitemShut {NoStop}%
\bibitem [{\citenamefont {Momenzadeh}\ \emph {et~al.}(2015)\citenamefont
  {Momenzadeh}, \citenamefont {St{\"o}hr}, \citenamefont {de~Oliveira},
  \citenamefont {Brunner}, \citenamefont {Denisenko}, \citenamefont {Yang},
  \citenamefont {Reinhard},\ and\ \citenamefont {Wrachtrup}}]{Ali_Pillar}%
  \BibitemOpen
  \bibfield  {author} {\bibinfo {author} {\bibfnamefont {S.~A.}\ \bibnamefont
  {Momenzadeh}}, \bibinfo {author} {\bibfnamefont {R.~J.}\ \bibnamefont
  {St{\"o}hr}}, \bibinfo {author} {\bibfnamefont {F.~F.}\ \bibnamefont
  {de~Oliveira}}, \bibinfo {author} {\bibfnamefont {A.}~\bibnamefont
  {Brunner}}, \bibinfo {author} {\bibfnamefont {A.}~\bibnamefont {Denisenko}},
  \bibinfo {author} {\bibfnamefont {S.}~\bibnamefont {Yang}}, \bibinfo {author}
  {\bibfnamefont {F.}~\bibnamefont {Reinhard}},\ and\ \bibinfo {author}
  {\bibfnamefont {J.}~\bibnamefont {Wrachtrup}},\ }\bibfield  {title} {\bibinfo
  {title} {Nanoengineered diamond waveguide as a robust bright platform for
  nanomagnetometry using shallow nitrogen vacancy centers},\ }\href
  {https://doi.org/10.1021/nl503326t} {\bibfield  {journal} {\bibinfo
  {journal} {Nano Letters}\ }\textbf {\bibinfo {volume} {15}},\ \bibinfo
  {pages} {165} (\bibinfo {year} {2015})}\BibitemShut {NoStop}%
\bibitem [{\citenamefont {Morigi}\ \emph {et~al.}(2000)\citenamefont {Morigi},
  \citenamefont {Eschner},\ and\ \citenamefont {Keitel}}]{EIT2}%
  \BibitemOpen
  \bibfield  {author} {\bibinfo {author} {\bibfnamefont {G.}~\bibnamefont
  {Morigi}}, \bibinfo {author} {\bibfnamefont {J.}~\bibnamefont {Eschner}},\
  and\ \bibinfo {author} {\bibfnamefont {C.~H.}\ \bibnamefont {Keitel}},\
  }\bibfield  {title} {\bibinfo {title} {Ground state laser cooling using
  electromagnetically induced transparency},\ }\href
  {https://doi.org/10.1103/PhysRevLett.85.4458} {\bibfield  {journal} {\bibinfo
   {journal} {Phys. Rev. Lett.}\ }\textbf {\bibinfo {volume} {85}},\ \bibinfo
  {pages} {4458} (\bibinfo {year} {2000})}\BibitemShut {NoStop}%
\bibitem [{\citenamefont {Khanin}\ and\ \citenamefont
  {Kocharovskaya}(1990)}]{CPT2}%
  \BibitemOpen
  \bibfield  {author} {\bibinfo {author} {\bibfnamefont {Y.~I.}\ \bibnamefont
  {Khanin}}\ and\ \bibinfo {author} {\bibfnamefont {O.~A.}\ \bibnamefont
  {Kocharovskaya}},\ }\bibfield  {title} {\bibinfo {title} {Inversionless
  amplification of ultrashort pulses and coherent population trapping in a
  three-level medium},\ }\href {https://doi.org/10.1364/JOSAB.7.002016}
  {\bibfield  {journal} {\bibinfo  {journal} {J. Opt. Soc. Am. B}\ }\textbf
  {\bibinfo {volume} {7}},\ \bibinfo {pages} {2016} (\bibinfo {year}
  {1990})}\BibitemShut {NoStop}%
\bibitem [{\citenamefont {Arimondo}\ and\ \citenamefont
  {Orriols}(1976)}]{CPT3}%
  \BibitemOpen
  \bibfield  {author} {\bibinfo {author} {\bibfnamefont {E.}~\bibnamefont
  {Arimondo}}\ and\ \bibinfo {author} {\bibfnamefont {G.}~\bibnamefont
  {Orriols}},\ }\bibfield  {title} {\bibinfo {title} {Nonabsorbing atomic
  coherences by coherent two-photon transitions in a three-level optical
  pumping},\ }\href@noop {} {\bibfield  {journal} {\bibinfo  {journal}
  {Letter.e al nuovo cimento}\ }\textbf {\bibinfo {volume} {17}} (\bibinfo
  {year} {1976})}\BibitemShut {NoStop}%
\bibitem [{\citenamefont {Gray}\ \emph {et~al.}(1978)\citenamefont {Gray},
  \citenamefont {Whitley},\ and\ \citenamefont {Stroud}}]{CPT4}%
  \BibitemOpen
  \bibfield  {author} {\bibinfo {author} {\bibfnamefont {H.~R.}\ \bibnamefont
  {Gray}}, \bibinfo {author} {\bibfnamefont {R.~M.}\ \bibnamefont {Whitley}},\
  and\ \bibinfo {author} {\bibfnamefont {C.~R.}\ \bibnamefont {Stroud}},\
  }\bibfield  {title} {\bibinfo {title} {Coherent trapping of atomic
  populations},\ }\href {https://doi.org/10.1364/OL.3.000218} {\bibfield
  {journal} {\bibinfo  {journal} {Opt. Lett.}\ }\textbf {\bibinfo {volume}
  {3}},\ \bibinfo {pages} {218} (\bibinfo {year} {1978})}\BibitemShut {NoStop}%
\bibitem [{\citenamefont {Yoo}\ and\ \citenamefont {Eberly}(1985)}]{CPT5}%
  \BibitemOpen
  \bibfield  {author} {\bibinfo {author} {\bibfnamefont {H.-I.}\ \bibnamefont
  {Yoo}}\ and\ \bibinfo {author} {\bibfnamefont {J.}~\bibnamefont {Eberly}},\
  }\bibfield  {title} {\bibinfo {title} {Dynamical theory of an atom with two
  or three levels interacting with quantized cavity fields},\ }\href
  {https://doi.org/https://doi.org/10.1016/0370-1573(85)90015-8} {\bibfield
  {journal} {\bibinfo  {journal} {Physics Reports}\ }\textbf {\bibinfo {volume}
  {118}},\ \bibinfo {pages} {239} (\bibinfo {year} {1985})}\BibitemShut
  {NoStop}%
\bibitem [{\citenamefont {Brewer}\ and\ \citenamefont {Hahn}(1975)}]{CPT6}%
  \BibitemOpen
  \bibfield  {author} {\bibinfo {author} {\bibfnamefont {R.~G.}\ \bibnamefont
  {Brewer}}\ and\ \bibinfo {author} {\bibfnamefont {E.~L.}\ \bibnamefont
  {Hahn}},\ }\bibfield  {title} {\bibinfo {title} {Coherent two-photon
  processes: Transient and steady-state cases},\ }\href
  {https://doi.org/10.1103/PhysRevA.11.1641} {\bibfield  {journal} {\bibinfo
  {journal} {Phys. Rev. A}\ }\textbf {\bibinfo {volume} {11}},\ \bibinfo
  {pages} {1641} (\bibinfo {year} {1975})}\BibitemShut {NoStop}%
\bibitem [{\citenamefont {Whitley}\ and\ \citenamefont {Stroud}(1976)}]{CPT7}%
  \BibitemOpen
  \bibfield  {author} {\bibinfo {author} {\bibfnamefont {R.~M.}\ \bibnamefont
  {Whitley}}\ and\ \bibinfo {author} {\bibfnamefont {C.~R.}\ \bibnamefont
  {Stroud}},\ }\bibfield  {title} {\bibinfo {title} {Double optical
  resonance},\ }\href {https://doi.org/10.1103/PhysRevA.14.1498} {\bibfield
  {journal} {\bibinfo  {journal} {Phys. Rev. A}\ }\textbf {\bibinfo {volume}
  {14}},\ \bibinfo {pages} {1498} (\bibinfo {year} {1976})}\BibitemShut
  {NoStop}%
\end{thebibliography}%


\begin{thebibliography}{8}%
\makeatletter
\providecommand \@ifxundefined [1]{%
 \@ifx{#1\undefined}
}%
\providecommand \@ifnum [1]{%
 \ifnum #1\expandafter \@firstoftwo
 \else \expandafter \@secondoftwo
 \fi
}%
\providecommand \@ifx [1]{%
 \ifx #1\expandafter \@firstoftwo
 \else \expandafter \@secondoftwo
 \fi
}%
\providecommand \natexlab [1]{#1}%
\providecommand \enquote  [1]{``#1''}%
\providecommand \bibnamefont  [1]{#1}%
\providecommand \bibfnamefont [1]{#1}%
\providecommand \citenamefont [1]{#1}%
\providecommand \href@noop [0]{\@secondoftwo}%
\providecommand \href [0]{\begingroup \@sanitize@url \@href}%
\providecommand \@href[1]{\@@startlink{#1}\@@href}%
\providecommand \@@href[1]{\endgroup#1\@@endlink}%
\providecommand \@sanitize@url [0]{\catcode `\\12\catcode `\$12\catcode
  `\&12\catcode `\#12\catcode `\^12\catcode `\_12\catcode `\%12\relax}%
\providecommand \@@startlink[1]{}%
\providecommand \@@endlink[0]{}%
\providecommand \url  [0]{\begingroup\@sanitize@url \@url }%
\providecommand \@url [1]{\endgroup\@href {#1}{\urlprefix }}%
\providecommand \urlprefix  [0]{URL }%
\providecommand \Eprint [0]{\href }%
\providecommand \doibase [0]{https://doi.org/}%
\providecommand \selectlanguage [0]{\@gobble}%
\providecommand \bibinfo  [0]{\@secondoftwo}%
\providecommand \bibfield  [0]{\@secondoftwo}%
\providecommand \translation [1]{[#1]}%
\providecommand \BibitemOpen [0]{}%
\providecommand \bibitemStop [0]{}%
\providecommand \bibitemNoStop [0]{.\EOS\space}%
\providecommand \EOS [0]{\spacefactor3000\relax}%
\providecommand \BibitemShut  [1]{\csname bibitem#1\endcsname}%
\let\auto@bib@innerbib\@empty
%</preamble>
\bibitem [{\citenamefont {Foglszinger}\ \emph {et~al.}(2022)\citenamefont
  {Foglszinger}, \citenamefont {Denisenko}, \citenamefont {Kornher},
  \citenamefont {Schreck}, \citenamefont {Knolle}, \citenamefont {Yavkin},
  \citenamefont {Kolesov},\ and\ \citenamefont {Wrachtrup}}]{Foglszinger2022}%
  \BibitemOpen
  \bibfield  {author} {\bibinfo {author} {\bibfnamefont {J.}~\bibnamefont
  {Foglszinger}}, \bibinfo {author} {\bibfnamefont {A.}~\bibnamefont
  {Denisenko}}, \bibinfo {author} {\bibfnamefont {T.}~\bibnamefont {Kornher}},
  \bibinfo {author} {\bibfnamefont {M.}~\bibnamefont {Schreck}}, \bibinfo
  {author} {\bibfnamefont {W.}~\bibnamefont {Knolle}}, \bibinfo {author}
  {\bibfnamefont {B.}~\bibnamefont {Yavkin}}, \bibinfo {author} {\bibfnamefont
  {R.}~\bibnamefont {Kolesov}},\ and\ \bibinfo {author} {\bibfnamefont
  {J.}~\bibnamefont {Wrachtrup}},\ }\bibfield  {title} {\bibinfo {title} {Tr12
  centers in diamond as a room temperature atomic scale vector magnetometer},\
  }\href {https://doi.org/10.1038/s41534-022-00566-8} {\bibfield  {journal}
  {\bibinfo  {journal} {npj Quantum Information}\ }\textbf {\bibinfo {volume}
  {8}},\ \bibinfo {pages} {65} (\bibinfo {year} {2022})}\BibitemShut {NoStop}%
\bibitem [{\citenamefont {Khanin}\ and\ \citenamefont
  {Kocharovskaya}(1990)}]{CPT2}%
  \BibitemOpen
  \bibfield  {author} {\bibinfo {author} {\bibfnamefont {Y.~I.}\ \bibnamefont
  {Khanin}}\ and\ \bibinfo {author} {\bibfnamefont {O.~A.}\ \bibnamefont
  {Kocharovskaya}},\ }\bibfield  {title} {\bibinfo {title} {Inversionless
  amplification of ultrashort pulses and coherent population trapping in a
  three-level medium},\ }\href {https://doi.org/10.1364/JOSAB.7.002016}
  {\bibfield  {journal} {\bibinfo  {journal} {J. Opt. Soc. Am. B}\ }\textbf
  {\bibinfo {volume} {7}},\ \bibinfo {pages} {2016} (\bibinfo {year}
  {1990})}\BibitemShut {NoStop}%
\bibitem [{\citenamefont {Arimondo}\ and\ \citenamefont
  {Orriols}(1976)}]{CPT3}%
  \BibitemOpen
  \bibfield  {author} {\bibinfo {author} {\bibfnamefont {E.}~\bibnamefont
  {Arimondo}}\ and\ \bibinfo {author} {\bibfnamefont {G.}~\bibnamefont
  {Orriols}},\ }\bibfield  {title} {\bibinfo {title} {Nonabsorbing atomic
  coherences by coherent two-photon transitions in a three-level optical
  pumping},\ }\href@noop {} {\bibfield  {journal} {\bibinfo  {journal}
  {Letter.e al nuovo cimento}\ }\textbf {\bibinfo {volume} {17}} (\bibinfo
  {year} {1976})}\BibitemShut {NoStop}%
\bibitem [{\citenamefont {Gray}\ \emph {et~al.}(1978)\citenamefont {Gray},
  \citenamefont {Whitley},\ and\ \citenamefont {Stroud}}]{CPT4}%
  \BibitemOpen
  \bibfield  {author} {\bibinfo {author} {\bibfnamefont {H.~R.}\ \bibnamefont
  {Gray}}, \bibinfo {author} {\bibfnamefont {R.~M.}\ \bibnamefont {Whitley}},\
  and\ \bibinfo {author} {\bibfnamefont {C.~R.}\ \bibnamefont {Stroud}},\
  }\bibfield  {title} {\bibinfo {title} {Coherent trapping of atomic
  populations},\ }\href {https://doi.org/10.1364/OL.3.000218} {\bibfield
  {journal} {\bibinfo  {journal} {Opt. Lett.}\ }\textbf {\bibinfo {volume}
  {3}},\ \bibinfo {pages} {218} (\bibinfo {year} {1978})}\BibitemShut {NoStop}%
\bibitem [{\citenamefont {Yoo}\ and\ \citenamefont {Eberly}(1985)}]{CPT5}%
  \BibitemOpen
  \bibfield  {author} {\bibinfo {author} {\bibfnamefont {H.-I.}\ \bibnamefont
  {Yoo}}\ and\ \bibinfo {author} {\bibfnamefont {J.}~\bibnamefont {Eberly}},\
  }\bibfield  {title} {\bibinfo {title} {Dynamical theory of an atom with two
  or three levels interacting with quantized cavity fields},\ }\href
  {https://doi.org/https://doi.org/10.1016/0370-1573(85)90015-8} {\bibfield
  {journal} {\bibinfo  {journal} {Physics Reports}\ }\textbf {\bibinfo {volume}
  {118}},\ \bibinfo {pages} {239} (\bibinfo {year} {1985})}\BibitemShut
  {NoStop}%
\bibitem [{\citenamefont {Brewer}\ and\ \citenamefont {Hahn}(1975)}]{CPT6}%
  \BibitemOpen
  \bibfield  {author} {\bibinfo {author} {\bibfnamefont {R.~G.}\ \bibnamefont
  {Brewer}}\ and\ \bibinfo {author} {\bibfnamefont {E.~L.}\ \bibnamefont
  {Hahn}},\ }\bibfield  {title} {\bibinfo {title} {Coherent two-photon
  processes: Transient and steady-state cases},\ }\href
  {https://doi.org/10.1103/PhysRevA.11.1641} {\bibfield  {journal} {\bibinfo
  {journal} {Phys. Rev. A}\ }\textbf {\bibinfo {volume} {11}},\ \bibinfo
  {pages} {1641} (\bibinfo {year} {1975})}\BibitemShut {NoStop}%
\bibitem [{\citenamefont {Whitley}\ and\ \citenamefont {Stroud}(1976)}]{CPT7}%
  \BibitemOpen
  \bibfield  {author} {\bibinfo {author} {\bibfnamefont {R.~M.}\ \bibnamefont
  {Whitley}}\ and\ \bibinfo {author} {\bibfnamefont {C.~R.}\ \bibnamefont
  {Stroud}},\ }\bibfield  {title} {\bibinfo {title} {Double optical
  resonance},\ }\href {https://doi.org/10.1103/PhysRevA.14.1498} {\bibfield
  {journal} {\bibinfo  {journal} {Phys. Rev. A}\ }\textbf {\bibinfo {volume}
  {14}},\ \bibinfo {pages} {1498} (\bibinfo {year} {1976})}\BibitemShut
  {NoStop}%
\bibitem [{\citenamefont {Balasubramanian}\ \emph {et~al.}(2008)\citenamefont
  {Balasubramanian}, \citenamefont {Chan}, \citenamefont {Kolesov},
  \citenamefont {Al-Hmoud}, \citenamefont {Tisler}, \citenamefont {Shin},
  \citenamefont {Kim}, \citenamefont {Wojcik}, \citenamefont {Hemmer},
  \citenamefont {Krueger}, \citenamefont {Hanke}, \citenamefont
  {Leitenstorfer}, \citenamefont {Bratschitsch}, \citenamefont {Jelezko},\ and\
  \citenamefont {Wrachtrup}}]{NV_sense_mag}%
  \BibitemOpen
  \bibfield  {author} {\bibinfo {author} {\bibfnamefont {G.}~\bibnamefont
  {Balasubramanian}}, \bibinfo {author} {\bibfnamefont {I.~Y.}\ \bibnamefont
  {Chan}}, \bibinfo {author} {\bibfnamefont {R.}~\bibnamefont {Kolesov}},
  \bibinfo {author} {\bibfnamefont {M.}~\bibnamefont {Al-Hmoud}}, \bibinfo
  {author} {\bibfnamefont {J.}~\bibnamefont {Tisler}}, \bibinfo {author}
  {\bibfnamefont {C.}~\bibnamefont {Shin}}, \bibinfo {author} {\bibfnamefont
  {C.}~\bibnamefont {Kim}}, \bibinfo {author} {\bibfnamefont {A.}~\bibnamefont
  {Wojcik}}, \bibinfo {author} {\bibfnamefont {P.~R.}\ \bibnamefont {Hemmer}},
  \bibinfo {author} {\bibfnamefont {A.}~\bibnamefont {Krueger}}, \bibinfo
  {author} {\bibfnamefont {T.}~\bibnamefont {Hanke}}, \bibinfo {author}
  {\bibfnamefont {A.}~\bibnamefont {Leitenstorfer}}, \bibinfo {author}
  {\bibfnamefont {R.}~\bibnamefont {Bratschitsch}}, \bibinfo {author}
  {\bibfnamefont {F.}~\bibnamefont {Jelezko}},\ and\ \bibinfo {author}
  {\bibfnamefont {J.}~\bibnamefont {Wrachtrup}},\ }\bibfield  {title} {\bibinfo
  {title} {Nanoscale imaging magnetometry with diamond spins under ambient
  conditions},\ }\href {https://doi.org/10.1038/nature07278} {\bibfield
  {journal} {\bibinfo  {journal} {Nature}\ }\textbf {\bibinfo {volume} {455}},\
  \bibinfo {pages} {648} (\bibinfo {year} {2008})}\BibitemShut {NoStop}%
\end{thebibliography}%
\end{document}

%EMAILS HAVE TO BE HERE!!!

\preprint{APS/123-QED}

\title{Discovery of ST2 centers in natural and CVD diamond}% Force line breaks with \\

\author{Jonas Foglszinger$^1$}
\author{Andrej Denisenko$^1$}%
\author{Georgy V. Astakhov$^2$}
\author{Lev Kazak$^3$}
\author{Petr Siyushev$^4$}
\author{Alexander M. Zaitsev$^5$}
\author{Roman Kolesov$^1$}%
\author{J\"org Wrachtrup$^1$}%
\affiliation{$^1$3rd Institute of Physics, University of Stuttgart,70569 Stuttgart, Germany}%
\affiliation{$^2$Helmholtz-Zentrum Dresden-Rossendorf,
Institute of Ion Beam Physics and Materials Research,
01328 Dresden, Germany}
\affiliation{$^3$University Institute for Quantum Optics, Ulm University, Albert-Einstein-Allee 11, D-89081 Ulm, Germany}
\affiliation{$^4$Institute for Materials Research, Hasselt University, Wetenschapspark 1, 3590 Diepenbeek, Belgium}
\affiliation{$^5$ Missing affiliation}
\date{\today}% It is always \today, today,

\maketitle
\section*{\label{supp0}SUPPLEMENTARY NOTE 1: Preparation of diamond samples}
All artificial diamond samples used in this study are exclusively supplied by \emph{Element Six (E6) Technologies}.
These diamonds are single crystals, grown through chemical vapor deposition (CVD).
We select from their \emph{Quantum / Radiation Detectors} portfolio, ensuring access to the highest quality material available.
This selection is driven by the need for diamonds with minimal intrinsic strain and negligible fluorescent background, critical for the highly sensitive experiments conducted.
The emphasis on low background fluorescence is particularly important, as the experiments predominantly use \SI{410}{\nano\metre} laser excitation, a short wavelength known to excite more fluorescence compared to the commonly used \SI{532}{\nano\metre} excitation for NV center studies.
The ST2 centers under investigation exhibit relatively low fluorescence, producing only about 10 to \SI{20}{kcps}.
This underscores the necessity of minimizing background noise to maintain an acceptable signal-to-noise ratio for all subsequent experiments.
For those trying to replicate these results, we caution that using lower-quality diamonds will likely result in failure to detect these centers.
For attempts to replicate these experiments, diamonds of the same or comparable quality from reliable suppliers are almost certainly mandatory. 
In this study, we specifically use highly pure Type II diamonds, characterized by nitrogen concentrations below \SI{5}{ppb} and boron concentrations below \SI{1}{ppb}.
This choice is based on the first successful attempt to produce ST2 centers artificially, which happened to occur in a diamond with these specific configurations.
However, these conditions are not confirmed to be mandatory. In fact, in a later attempt, ST2 centers were produced within a boron-doped diamond layer, showing no observable differences. The connection between ST2 centers and nitrogen remains uncharted. Nonetheless, the negative correlation between nitrogen and the related TR12 defect suggests the possibility of some level of interference \cite{Foglszinger2022}.
\section*{\label{supp1}SUPPLEMENTARY NOTE 2: Creation of ST2 centers}
As described in the main manuscript, the method to generate ST2 centers is straightforward. A freshly prepared diamond, as described in the previous section, is implanted with \SI{10}{keV} $^{12}$C ions. Implantation is performed into the (100) plane with an additional \SI{7}{\degree} tilt to prevent channeling effects. A dose of $2 \times 10^{14}$ ions/cm$^2$ or higher was used. Working with doses that surpass the graphitization threshold is also a viable approach.

No ST2 centers are observed immediately after implantation, nor does prolonged exposure to \SI{410}{\nano\metre} laser excitation activate them. This confirms that activation requires an additional annealing step, typically performed for \SI{1}{h}. 
The annealing temperature was tested between \SI{800}{\celsius} and \SI{1600}{\celsius}. 
Activation was found to require at least \SI{1100}{\celsius}, as shown in Figure 1c of the main manuscript.
Formation efficiency peaks between \SI{1200}{\celsius} and \SI{1300}{\celsius}, but drops off at temperatures above \SI{1400}{\celsius}.
However, the decline above \SI{1400}{\celsius} remains unclear, due to possible confounding factors.
At annealing temperatures of \SI{1400}{\celsius} and above, the surface of the diamond experiences slight graphitization, which effectively reduces the thickness of the ST2-containing layer. This may partially explain the reduction in ST2 centers without requiring a decrease in formation efficiency. Furthermore, the only sample annealed at \SI{1600}{\celsius} underwent a total of \SI{12}{h} of annealing, rather than the usual \SI{1}{h}, and was implanted with significantly higher energies (\SI{100}{keV}-\SI{500}{keV}). The potential effects of these variations remain unclear.

As outlined in the main manuscript, one test sample annealed at \SI{1200}{\celsius} underwent slow etching using reactive ion etching (RIE) with oxygen plasma. The etching rate was kept relatively low, around \SI{5}{\nano\metre / \minute}, as higher etching rates were found to significantly increase the formation of unintended centers, such as di-vacancies, complicating the evaluation process. 
The results reveal a clear linear relationship between the number of ST2 centers and the total number of vacancies produced by implantation, as simulated by SRIM, suggesting the intrinsic nature of the ST2 defect. Additionally, it is known %\comment{REF?} 
that vacancies can migrate deeper into the diamond during annealing. If vacancies alone were involved in the formation of ST2 centers, we would expect the centers to also be located deeper than the vacancy profile directly induced by implantation. The absence of such deeper ST2 centers suggests the involvement of interstitial carbon atoms in their formation.

In an effort to apply these findings, various methods to produce ST2 centers were tested. 
The absence of foreign elements in the initial attempts suggests that any approach capable of generating sufficient lattice damage in the diamond, without actively quenching the process, should lead to the formation of ST2 centers.
With this in mind, we attempt to produce ST2 centers using two methods: first, by implanting helium (He) ions at a dose creating lattice damage similar to that of the previously used carbon implants; and second, by implanting lead (Pb) ions at lower concentrations. 
Although the total damage from the Pb implantation is significantly lower, the localized damage within individual Pb implantation tracks should theoretically be high enough to form ST2 centers.
Both implantation procedures were followed by a 1-hour annealing step at \SI{1200}{\celsius}. However, no ST2 centers were observed.
More surprisingly, a subsequent re-implantation with $^{12}$C ions leads to the formation of ST2 centers in both cases, confirming that neither He nor Pb ions actively prevent the formation of ST2 centers. At this point, we can only speculate that an overabundance of interstitial carbon ions in the diamond might be directly required for the formation process. 

Although the method for creating ST2 centers is reproducible, it is relatively inefficient. From the data in Figure 1d of the main manuscript, an optimal dose for ST2 center formation can be derived.
The dose should be maximized, but must not exceed the threshold where amorphous diamond begins to form. In Figure 1d, areas of graphitized and amorphous diamond are marked. For a 10 keV $^{12}$C implantation, the threshold for amorphization is approximately 9.5 $\times$ 10$^{14}$ ions/cm$^2$. With a conversion rate of $2 \times 10^{-8}$ centers per ion for \SI{10}{keV} implantation, this results in approximately 120 ST2 centers in a \SI{10}{\micro\metre} $\times$ \SI{10}{\micro\metre} field.
\section*{\label{supp1}SUPPLEMENTARY NOTE 3:  Experimental setup}
The setup used for the experiments in this paper is an updated version of the one described in \cite{Foglszinger2022}.
As a result, certain similarities in the description are unavoidable. Since the data evaluation was also conducted in a similar manner, the theoretical description provided here will also show notable similarities.
\begin{figure}
     \centering
         \includegraphics[width=0.45\textwidth]{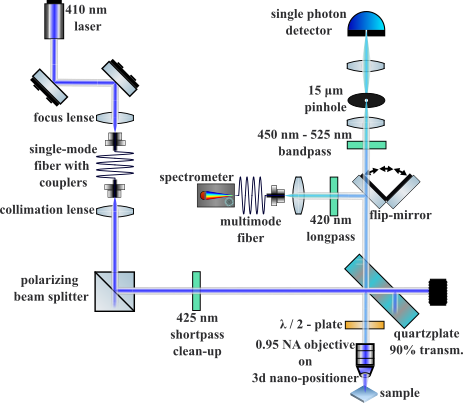}
     \caption{Schematic representations of the home-built confocal microscope system are provided, showcasing all relevant components.}
\end{figure}
The optical and magnetic characterizations are performed using a home-built confocal microscope, based on \SI{410}{\nano\metre} linearly polarized laser excitation (see Supplementary Figure 1), operating at room temperature. The \SI{410}{\nano\metre} laser light is coupled into a \SI{405}{\nano\metre} single-mode polarization-maintaining optical fiber, providing a well-defined Gaussian mode profile. This configuration also isolates the laser from the rest of the setup, making power adjustments easier without introducing errors. Although the polarization is generally well-defined at this stage, the beam is reflected at a polarizing beam splitter (PBS) to ensure reproducible polarization in case of maintenance.
The addition of a $\lambda$/2 plate just before the objective allows measurements that depend on the orientation of the laser’s linear polarization.
To reduce noise in the excitation path, a high-quality clean-up filter is installed.
At the hearth of the setup is a \SI{0.95}{NA} objective with a large back aperture of around \SI{50}{\milli\metre\squared}. 
This high NA objective was essential for delivering the high local laser power required to saturate ST2 centers. Its broad acceptance angle also allowed effective fluorescence collection.
Moreover, the objective was attached to a 3D nanopositioner, enabling precise control of the excitation point on the sample.
The fluorescence light, collected through the same objective, was separated from the excitation light using a quartz plate with \SI{10}{\percent} reflectivity and \SI{90}{\percent} transmission.
The fluorescence was then filtered through a bandpass filter covering \SI{450}{\nano\metre} to \SI{525}{\nano\metre} (\SI{2.76}{eV} to \SI{2.36}{eV}).
According to Figure 1b in the main manuscript, this range captures the majority of ST2 fluorescence while effectively suppressing Raman scattering.
Using even shorter excitation wavelengths might potentially further improve the signal-to-noise ratio, as this would allow complete removal of second-order Raman scattering without clipping the fluorescence signal from the centers. 
However, such an enhancement would place higher demands on the laser system, as \SI{410}{\nano\metre} technology is more commonly available.
The filtered fluorescence light is focused through a pinhole, ensuring the confocal nature of the setup.
Pinhole sizes of 10, 15, and \SI{20}{\micro\metre} were tested, with the \SI{15}{\micro\metre} pinhole delivering the optimal signal-to-noise ratio in our setup.
The remaining fluorescence is then directed to a single-photon detector based on the principle of an avalanche photodiode (APD).
The photon counts from this APD is recorded using a National Instruments NI card.
Alternatively, a flip-mirror can be inserted into the collection path before the bandpass, redirecting all fluorescence light into a secondary detection path leading to a spectrometer.
The light is then focused onto a multimode fiber with a \SI{25}{\micro\meter} core, which serves as an alternative pinhole, ensuring the setup remains confocal.  
The fiber is directly coupled to the spectrometer.

A permanent magnet generates the magnetic field applied to the sample.
The magnet’s position can be adjusted via stepper motors, which move independent of the sample holder and are fixed to the optics table, as explained in the main manuscript.
Additionally, the NI-card mentioned earlier manages all pulsed microwave sequences described in the main text. This is implemented through voltage-controlled switches placed between the microwave source and the sample.

\section*{SUPPLEMENTARY NOTE 4: Coherent population trapping }
Coherent population trapping (CPT) occurs in a three-level system when it is driven by two resonant electromagnetic fields. For this phenomenon to arise, two of the three states must have long lifetimes, while the third state is short-lived. A more detailed analysis of this effect can be found in several research papers such as \cite{CPT2,CPT3,CPT4, CPT5,CPT6,CPT7}.

Consider a three-level system, where the states are arbitrarily labeled $\ket{T_\text{x}}$, $\ket{T_\text{y}}$, and $\ket{T_\text{z}}$.
This labeling, though unconventional for the general case, was specifically chosen to reflect the occurrence of CPT (Figure 2d) in the ST2 level system shown in Figure 2a of the main manuscript.
The unperturbed Hamiltonian of this three-level system can be expressed as:
\begin{equation}
\label{CPT1}
\text{H}_\text{0}=\hbar\omega_\text{x}\ket{T_\text{x}}\bra{T_\text{x}}+\hbar\omega_\text{y}\ket{T_y}\bra{T_\text{y}}+\hbar\omega_\text{z}\ket{T_\text{z}}\bra{T_\text{z}}\,.
\end{equation}
So far, we have only defined the energy $\hbar\omega_i$ of each state $\ket{T_i}$.
For simplicity, we arbitrarily assign $\ket{T_\text{z}}$ as the short-lived state, while $\ket{T_\text{x}}$ and $\ket{T_\text{y}}$ are chosen as long-lived.
This choice is again purely to match the situation in the ST2 system.
To observe the effect of coherent population trapping (CPT), the system is now driven by two resonant microwave (MW) fields.
Each field driving the transition from the short-lived state $\ket{T_\text{z}}$ to one of the long-lived states, $\ket{T_\text{x}}$ or $\ket{T_\text{y}}$.
This is represented mathematically by introducing a perturbation term, $\text{H}_1$, to the unperturbed Hamiltonian $\text{H}_0$:
\begin{equation}
\text{H}_\text{1}=-\frac{\hbar}{2}\left(\Omega_\text{p}e^{-\text{i}\omega_\text{p}t}\ket{T_\text{x}}\bra{T_\text{z}}+\Omega_\text{c}\text{e}^{-\text{i}\omega_\text{c}t}\ket{T_\text{y}}\bra{T_\text{z}}\right) + \text{H.c}.
\end{equation}
$\Omega_\text{p}$ and $\Omega_\text{c}$ denote the two Rabi frequencies, associated with the transition energies $\omega_\text{p}=\omega_\text{x}-\omega_\text{z}$ and $\omega_\text{c}=\omega_\text{y}-\omega_\text{z}$.
So far, we have laid the groundwork but have not yet performed any actual calculations. To proceed, we need to solve the Schrödinger equation for the full Hamiltonian, $\text{H} = \text{H}_0 + \text{H}_1$. For this, we choose a general ansatz for the wavefunction, $\ket{\psi(t)}$:
\begin{equation}
\ket{\psi(t)}=c_\text{x}(t)e^{-\text{i}\omega_\text{x}t}\ket{T_\text{x}}+c_\text{y}(t)e^{-\text{i}\omega_\text{y}t}\ket{T_\text{y}}+c_\text{z}(t)e^{-\text{i}\omega_\text{z}t}\ket{T_\text{z}}\,.
\end{equation}
If this ansatz is inserted into the Schrödinger equation, it immediately leads to a simple set of differential equations for the coefficient functions $c_i$.
\begin{equation}
\begin{split}
\dot{c_\text{x}}&=\frac{\text{i}}{2}\Omega_\text{p}c_\text{z}\\
\dot{c_\text{y}}&=\frac{\text{i}}{2}\Omega_\text{c}c_\text{z}\\
\dot{c_\text{z}}&=\frac{\text{i}}{2}\left(\Omega_\text{p}c_\text{x}+\Omega_\text{c}c_\text{y}\right)\,.\\
\end{split}
\end{equation}
Instead of attempting to solve this differential equation in general, which is straightforward but beyond the scope of this brief discussion, we simply present the solution.
We neither prove that this is the sole solution, nor is that required for the argument that follows. The given solution to the set of differential equations is:
\begin{equation}
\begin{split}
c_\text{x}=\cos(\theta),\quad c_\text{y}=-\sin(\theta)\quad\text{ and }\quad c_\text{z}=0\quad \text{where}\\
\cos(\theta)=\frac{\Omega_\text{c}}{\sqrt{\Omega_\text{c}^2+\Omega_\text{p}^2}}\quad\text{ and }\quad 
\sine(\theta)=\frac{\Omega_\text{p}}{\sqrt{\Omega_\text{c}^2+\Omega_\text{p}^2}}\,,
\label{CPT2}
\end{split}
\end{equation}
We leave it to the reader to confirm that this is indeed a solution. From an interpretation standpoint, this indicates that such a system is capable of developing a dark state $\ket{\psi_\text{D}}=\cos(\theta)\cdot\ket{T_\text{x}}-\sin(\theta)\cdot\ket{T_\text{y}}$.
Despite both long-lived states being resonantly driven via microwaves to interact with the short-lived state, this dark state has no probability of being in or transitioning to $\ket{T_\text{z}}$.
When the system remains in this superposition of $\ket{T_\text{x}}$ and $\ket{T_\text{y}}$, the effect is called coherent population trapping.

Since only long-lived states are involved in this dark state by design, it inherits its long lifetime from $\ket{T_\text{x}}$ and $\ket{T_\text{y}}$.
This leads to observable phenomena associated with the effect having a relatively narrow natural linewidth.
In Figure 2d of the main manuscript, this corresponds to the narrow fluorescence dip with a width of approximately \SI{35}{KHz}.

In the experiment shown in Figure 2d, one microwave frequency $\omega_\text{pump}$ was parked on the transition frequency between $\ket{T_\text{z}}$ and $\ket{T_\text{x}}$, which corresponds to $\omega_p$ in the notation above.
A second microwave frequency, $\omega_\text{probe}$, was then swept from $\omega_c - \epsilon$ to $\omega_c + \epsilon$, thus crossing the transition frequency between $\ket{T_\text{z}}$ and $\ket{T_\text{y}}$. 
As the probe frequency approaches $\omega_c$, the fluorescence increases. This peak is a standard ODMR resonance and is not related to the effect of CPT.
The width of this peak is primarily determined by the lifetime of the short-lived state $\ket{T_\text{z}}$.
However, as the probe frequency approaches $\omega_c$ even more closely, a distinct phenomenon occurs: a narrow dip in fluorescence, embedded within the broader fluorescence peak.
This dip, whose width is correlated with the lifetime of the long-lived states $\ket{T_\text{x}}$ and $\ket{T_\text{y}}$, is the direct observation of coherent population trapping.  
Since this phenomenon can only occur when two long-lived states are present, it directly matches with the proposed model for ST2.
\section*{SUPPLEMENTARY NOTE 5: Non Equilibrium Steady state solution}
In order to properly understand the dynamics of the ST2 system, we need to recognize that we typically operate it in a so-called non-equilibrium steady state (NESS). 
This is due to the fact that the system is driven by a laser. Consequently, the system can never reach actual equilibrium.

However, having waited a sufficiently long period of time, the system stabilizes in a probabilistic manner such that, for any given point in time $t_0$, the probability of the system being found in each state does no longer change. 
This fixed probability distribution of the population is what we call the NESS of the system.

In order to derive the NESS, we have to consider the matrix equation that characterizes the system:
\begin{equation}
\dot{\rho}=\text{M}\rho
\label{steady_state}
\end{equation}
with the vector \\
$\rho=(\rho_\text{g}(S_0), \rho_\text{e}(S_1), \rho_\text{x},\rho_\text{y},\rho_\text{z})^{tr}$, representing the sates, and a Matrix
\begin{equation}
\text{M}=\left(\begin{array}{c c c c c} 
-P&\Gamma&\Gamma_\text{X}&\Gamma_\text{Y}&\Gamma_\text{Z}\\
P& -\Gamma-\gamma_\text{x}-\gamma_\text{y}-\gamma_\text{z}&0&0&0\\
0&\gamma_\text{x}&-\Gamma_\text{X}&0&0\\
0&\gamma_\text{y}&0&-\Gamma_\text{Y}&0\\
0&\gamma_\text{z}&0&0&-\Gamma_\text{Z}\\\end{array}\right)\,.
\label{steady_state_2}
\end{equation}
The symbols used here are equivalent to those in Figure 2a of the main manuscript.
As mentioned earlier, the NESS implies that the probability distribution no longer changes over time.
This can be expressed as $\dot{\rho}=0$, which corresponds to finding the eigenvector of the matrix $\text{M}$ to the eigenvalue $\text{EW}=0$.
This condition also ensures the conservation of population.
Additionally, it is important to note that most of these variables are not directly observable. The only measurable quantity is the fluorescence intensity during the transition $S_1\rightarrow{S_0}$.
Since the probability of observing this transition is directly proportional to the population in $\rho_\text{e}(S_1)$, this is the only simulations result that can be compared directly.
\section{SUPPLEMENTARY NOTE 6: Rabi oscillation}
In order to account for the effect of microwaves on the system, we need to incorporate Rabi oscillations into equation \ref{steady_state_2}.
This is done by adding off-diagonal elements. 
If we select the transition between $\ket{T_\text{z}}$ and $\ket{T_\text{x}}$ to be driven, the final result takes the form:
\begin{equation}
\begin{split}
\dot{\rho_\text{g}}&=-P\cdot\rho_\text{g}+\Gamma\cdot\rho_\text{e}+\Gamma_\text{x}\cdot\rho_\text{x}+\Gamma_\text{y}\cdot\rho_\text{y}+\Gamma_\text{z}\cdot\rho_\text{z}\\
\dot{\rho_\text{e}}&=P\cdot\rho_\text{g} -(\Gamma+\gamma_\text{x}+\gamma_\text{y}+\gamma_\text{z})\cdot \rho_\text{e}\\
\dot{\rho_\text{x}}&=\gamma_\text{x}\cdot\rho_\text{e}-\Gamma_\text{x}\cdot\rho_\text{x} + \frac{\text{i}\Omega}{2}(\sigma_\text{zx}-\sigma_\text{xz})\\
\dot{\rho_\text{y}}&=\gamma_\text{y}\cdot\rho_\text{e}-\Gamma_\text{y}\cdot\rho_\text{y}\\
\dot{\rho_\text{z}}&=\gamma_\text{z}\cdot\rho_\text{e}-\Gamma_\text{z}\cdot\rho_\text{z} + \frac{\text{i}\Omega}{2}(\sigma_\text{xz}-\sigma_\text{zx})\,\\
\dot{\sigma}_\text{xz}&=\frac{\text{i}\Omega}{2}(\rho_\text{z}-\rho_\text{x})-\sigma_\text{xz}\left(\frac{\Gamma_\text{x}}{2}+\frac{\Gamma_\text{z}}{2}\right)\\
\dot{\sigma}_\text{zx}&=\frac{\text{i}\Omega}{2}(\rho_\text{x}-\rho_\text{z})-\sigma_\text{zx}\left(\frac{\Gamma_\text{x}}{2}+\frac{\Gamma_\text{z}}{2}\right)\,.
\label{Rabi_sim}
\end{split}
\end{equation}
Additional microwave fields can easily be incorporated into the equation by adding further elements in a similar manner.
However, when simulating actual measurements, it is essential to also account for the waiting times.
For instance, when observing a Rabi oscillation, it is important to properly simulate the initialization into the NESS, the $\pi$-pulse with or without laser excitation, the waiting period without any laser light, and the final readout. 
The readout must be understood not as a single point, but as a quantity proportional to the average population in the excited state, while the system keeps evolving. 
The length of the readout window plays a crucial role and does influence both the simulation and experimental results.
\section*{\label{supp2}SUPPLEMENTARY NOTE 7: NV center as reference for magnetic maps}
When measuring the magnetic maps, as shown in Figure 3a of the main manuscript, it is unclear at first, how the positions on the map correspond to orientations within the diamond lattice. 
While it is obvious, that a variety of magnetic field strengths and orientations are measured, no clear reference point is available.
Although the magnetic field of the permanent magnet could be simulated, the accuracy of such simulations is limited without precise calibration, such as the exact distance between the magnet and the sample.
As a result, these simulations as a standalone lack the required precision to pinpoint specific orientations.
Instead, an approach was used, in which the well-known NV center serves as a reference. By comparing magnetic maps from both the NV and ST2 centers for the exact same magnetic field orientations, a reliable reference was established.
Since NV centers fluoresnce is quenched for magnetic fields misaligned with [111] in diamond, this orientation could be identified with high precision. This revealed that the z-axis of ST2 centers is also aligned with [111] in diamond, with a precision of about \SI{5}{\degree}.
\section*{\label{supp3} SUPPLEMENTARY NOTE 8: Simulating magnetic maps and orientations}
From the previous section, we know that the $z$-orientation of the ST2 metastable triplet aligns with the [111] orientation of diamond.
Considering the four distinct orientations of [111] in diamond, combined with the three-fold rotational symmetry of the diamond lattice, this immediately suggests the existence of twelve differently oriented ST2 centers, which are organized into four triples.

The first step in simulating magnetic maps is to simulate the magnetic field induced by the used permanent magnet.
The magnet itself is therefore approximated as a cube made up of equally spaced magnetic dipoles. 
This simulated magnet is then virtually moved over the sample, reflecting the situation in the real experimental setup.
For each position, the magnetic field at the center of the sample is calculated, yielding the magnetic field in the lab frame. 
To transition into the system frame of the ST2 center, one of the twelve orientations hast to be chosen first.
Transitioning to the chosen system frame is then accomplished with a straightforward 3D rotation.

To complete the simulation, we now need to  simulate the impact of the found magnetic field on an ST2 center.
Since singlet states do not have a magnetic moment, only the triplet state will be affected by the magnetic field.
We therefore start by considering the unperturbed Hamiltonian of only the triplet state
\begin{equation}
\text{H}_0=D(\text{S}_\text{z}^2-S(S+1)/3) + E(\text{S}_\text{x}^2-\text{S}_\text{y}^2)
\label{SimMag1}
\end{equation}
characterized by the zero-field splitting parameters $D$ and $E$. 
$\text{S}_\text{x}$, $\text{S}_\text{y}$ and $\text{S}_\text{z}$  are the spin matrices for a spin 1 ($S=1$) system.
The eigenvectors of this Hamiltonian can be found to be $\ket{T_\text{z}}=(0,1,0)^{tr}$, $\ket{T_\text{x}}=(-1,0,1)^{tr}$ and $\ket{T_\text{y}}=(1,0,1)^{tr}$ with eigenvalues $\text{EW}_z=-2D/3$, $\text{EW}_x=D/3-E$ and $\text{EW}_y=D/3+E$. 
By defining $\text{EW}_i=\hbar\omega_i$, this Hamiltonian can be written in the same way as in the previous section, shown in equation \eqref{CPT1}.

When adding a static magnetic field, similar to the addition of microwaves, an additional perturbation term must be included in the Hamiltonian.
This perturbation, which is a Zeeman interaction, reads H$_1=g\mu_B \textbf{S}\cdot\textbf{B}$.
The complete Hamiltonian $\text{H}=\text{H}_0+\text{H}_1$ then reads
\begin{equation}
\begin{split}
\text{H}=&D(\text{S}_\text{z}^2-S(S+1)/3)\\
& + E(\text{S}_\text{x}^2-\text{S}_\text{y}^2) + g\mu_B \textbf{S}\cdot\textbf{B} ,
\end{split}
\end{equation}

with the assumption of an electron g-factor $g=2$ and the Bohr magneton $\mu_B$.
This Hamiltonian again provides three eigenvectors $\ket{\varphi_i}$.
However, these eigenvectors are now a function of the magnetic field  \textbf{B}.
For further analysis, it is beneficial to express these new eigenvectors as linear combinations of the unperturbed eigenvectors $\ket{T_i}$ derived from equation \eqref{SimMag1}.

\begin{equation}
\ket{\varphi_i}=\alpha_i\cdot \ket{T_\text{x}}+\beta_i\cdot\ket{T_\text{y}}+\zeta_i\cdot\ket{T_\text{z}}\,.
\end{equation}
$\alpha_i$, $\beta_i$ and $\zeta_i$  are then referred to as the mixing coefficients.

At this point, certain assumptions must be made regarding how this influences the dynamics of the system. 
 Returning to the unperturbed situation, the transition from the excited state to the metastable triplet is driven by some interaction Hamiltonian $\text{H}_{\text{int}}^\text{eT}$.
Similarly, there is an interaction Hamiltonian for the transition from the metastable triplet to the ground state$\text{H}_{\text{int}}^\text{Tg}$.
These interaction Hamiltonians remain unknown.
Only the transition rates out of the metastable triplet, $\Gamma_i=|\bra{g}\text{H}_\text{int}^\text{Tg}\ket{T_i}|^2$, are measured directly.
The rates into the metastable triplet $\gamma_i=|\bra{e}\text{H}_\text{int}^\text{eT}\ket{T_i}|^2$ can only be estimated at this stage.
The key assumption here is that all cross terms average to zero ($\bra{g}\text{H}_\text{int}^\text{Tg}\ket{T_i}\bra{T_j}\text{H}_\text{int}^\text{Tg}\ket{g}=\bra{e}\text{H}_\text{int}^\text{eT}\ket{T_i}\bra{T_j}\text{H}_\text{int}^\text{eT}\ket{e}=0$ for $i\neq j$).
With this assumption, the new transition rates $ \gamma_i' $ and $ \Gamma_i' $, into and out of the metastable triplet, can be calculated:
\begin{equation}
\begin{split}
\gamma_i'=&\left|\bra{e}\text{H}_\text{int}^\text{eT}\ket{\varphi_i}\right|^2\\
=&\left|\bra{e}\text{H}_\text{int}^\text{eT}\ket{\alpha_i\cdot T_\text{x}}\right|^2\\
&+\left|\bra{e}\text{H}_\text{int}^\text{eT}\ket{\beta_i \cdot T_\text{y}}\right|^2\\
&+\left|\bra{e}\text{H}_\text{int}^\text{eT}\ket{\zeta_i\cdot T_\text{z}}\right|^2\\
=&|\alpha_i|^2\cdot\gamma_\text{x}+|\beta_i|^2\cdot\gamma_\text{y}+|\zeta_i|^2\cdot\gamma_\text{z}\,
\label{gammaformel1}
\end{split}
\end{equation}
and
\begin{equation}
\begin{split}
\Gamma_i'=&\left|\bra{g}\text{H}_\text{int}^\text{Tg}\ket{\varphi_i}\right|^2\\
=&\left|\bra{g}\text{H}_\text{int}^\text{Tg}\ket{\alpha_i\cdot T_x}\right|^2\\
&+\left|\bra{g}\text{H}_\text{int}^\text{Tg}\ket{\beta_i \cdot T_\text{y}}\right|^2\\
&+\left|\bra{g}\text{H}_\text{int}^\text{Tg}\ket{\zeta_i\cdot T_\text{z}}\right|^2\\
=&|\alpha_i|^2\cdot\Gamma_\text{x}+|\beta_i|^2\cdot\Gamma_\text{y}+|\zeta_i|^2\cdot\Gamma_\text{z}\,.
\label{gammaformel2}
\end{split}
\end{equation}
The new transition rates now depend on both, the magnetic field strength and its orientation. As a result, for each configuration of the magnetic field, a different set of transition rates is derived, resulting in a different steady-state solution.

This leaves one degree of freedom. 
While the $z$-orientation of ST2 was determined by the comparison with the NV centers, the orientations of the $x$- and $y$-axes still are to be determined.
This is achieved by rotating the $x$- and $y$-axes around the $z$-orientation, ensuring they always form an orthogonal system. 
The simulation results are then fitted to the actual measurements, determining the correct rotation angle.
In this process, the $y$-axis of ST2 was found to be situated in the plane formed by two adjacent $\sigma$-bonds.

By applying additional simple 3D rotations, all twelve different magnetic maps can be simulated.
The results are shown in Supplementary Figure 2.
For comparison, examples of all the measured maps are also listed in the same figure below.

The fact that the simulations and measurements match so closely confirms the accuracy of the proposed electronic level structure for ST2 (Figure 2a of the main manuscript).
\section*{SUPPLEMENTARY NOTE 9: Simulating ODMR contrast}
Given that the measurement of magnetic fields under ambient conditions is one of the main potential applications for ST2, it is worthwhile to investigate the exact dependence of the ODMR contrast on the magnetic field.
In a nutshell, the simulations of Rabi oscillations and the static magnetic field need to be combined.
We start from equation \eqref{steady_state_2}, with the modified transition rates $\gamma_i \rightarrow \gamma_i'$ and $\Gamma_i \rightarrow \Gamma_i'$, which depend on the static magnetic field \textbf{B}.

Since we are interested  in CW-ODMR measurements, and not in the coherent properties of Rabi oscillations, we account for the effect of microwaves by simply adding additional transition rates between two of the triplet sublevels $\ket{\varphi_i}$ and $\ket{\varphi_j}$ (equation \eqref{steady_state_mag} for ${i,j}={1,2}$).

\begin{widetext}
\begin{equation}
\text{M}=\left(\begin{array}{c c c c c} 
-P&\Gamma&\Gamma_1'&\Gamma_2'&\Gamma_3'\\
P& -\Gamma-\gamma_1'-\gamma_2'-\gamma_3'&0&0&0\\
0&\gamma_1'&-\Gamma_1'-MW&MW&0\\
0&\gamma_2'&MW&-\Gamma_2'-MW&0\\
0&\gamma_3'&0&0&-\Gamma_3'\\\end{array}\right)
\label{steady_state_mag}
\end{equation}
\end{widetext}
Finding the eigenvector of this matrix corresponding to the eigenvalue $\text{EW}=0$ again provides the steady state solution. 
As a result of all these modifications, the fluorescence intensity, which is proportional to the population of the excited state $\rho_\text{e}$, becomes a function of the microwave intensity ($\rho_\text{e}$ = $\rho_\text{e}(MW)$).
For sufficiently large values of $MW$ this value saturates.
The ODMR contrast for an ST2 centers transition, denoted as $\eta$, is then defined as
\begin{equation}
\eta = \lim_{MW \to \infty} \frac{\rho_\text{e}(MW) - \rho_\text{e}(MW=0)}{\rho_\text{e}(MW=0)}\,.
\end{equation}
This definition is, what we refer to as the ODMR contrast in both the main manuscript and the supplementary sections.

To use ODMR measurements for reversely calculating the magnetic field strength, the frequencies of two transitions are required  \cite{NV_sense_mag}.
Given that a chain is only as strong as its weakest link, the effective ODMR contrast of an ST2 center in such a practical measurement is given by the second-highest ODMR contrast of each individual transition.
This interpretation is the basis for the ODMR contrast displayed in Figure 3d of the main manuscript. 
Unfortunately, the exact orientation of the magnetic field cannot be derived from ODMR measurements on a single center. 
Instead only a unique combination $\Delta=D\cos(2\theta) + 2E\cos(2\phi)\sin^2(\theta)$ can be extracted \cite{NV_sense_mag}. 
\begin{figure*}
     \centering
         \includegraphics[width=\textwidth]{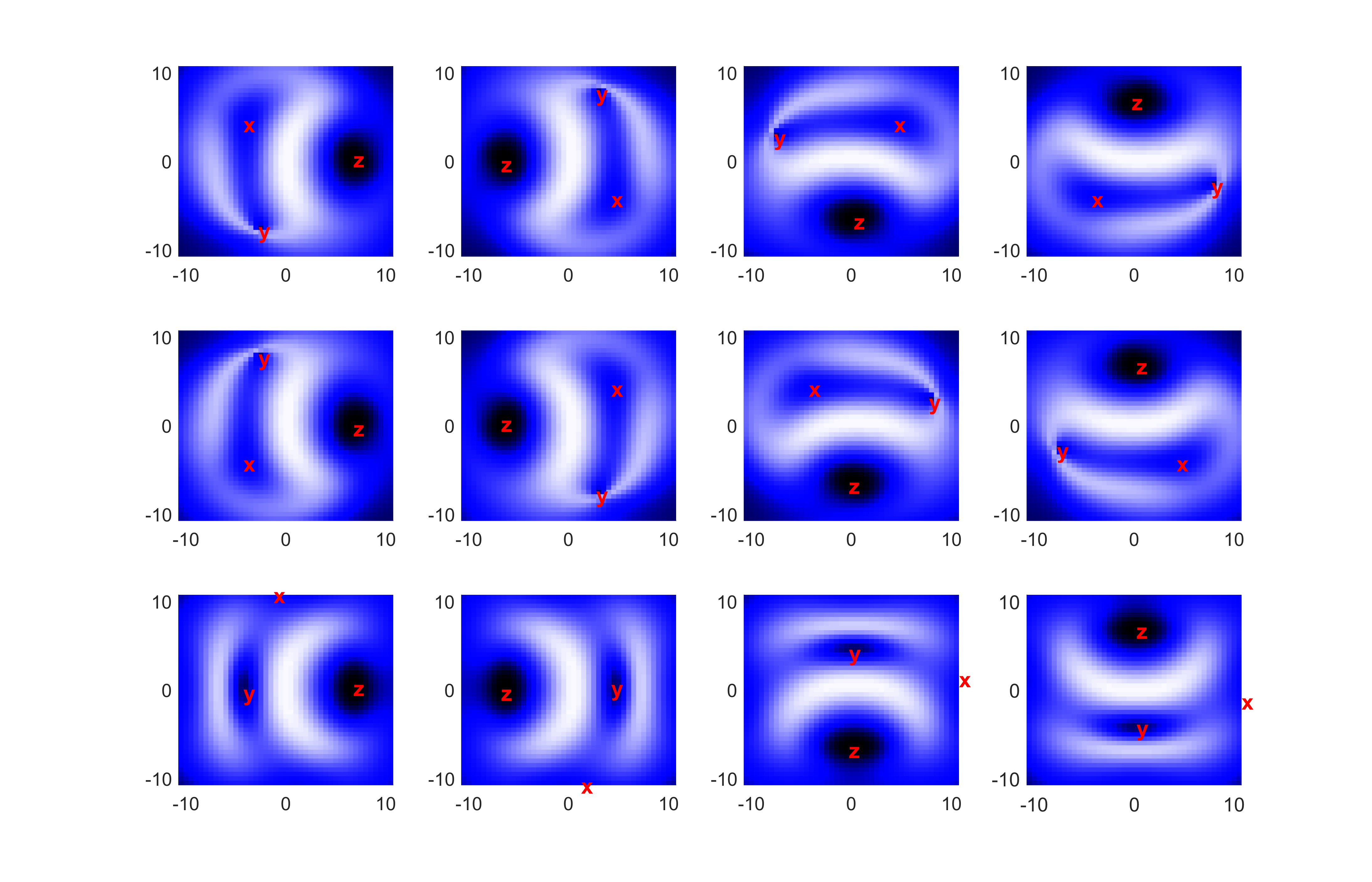}
		%\vspace{15pt}\par
          \includegraphics[width=\textwidth]{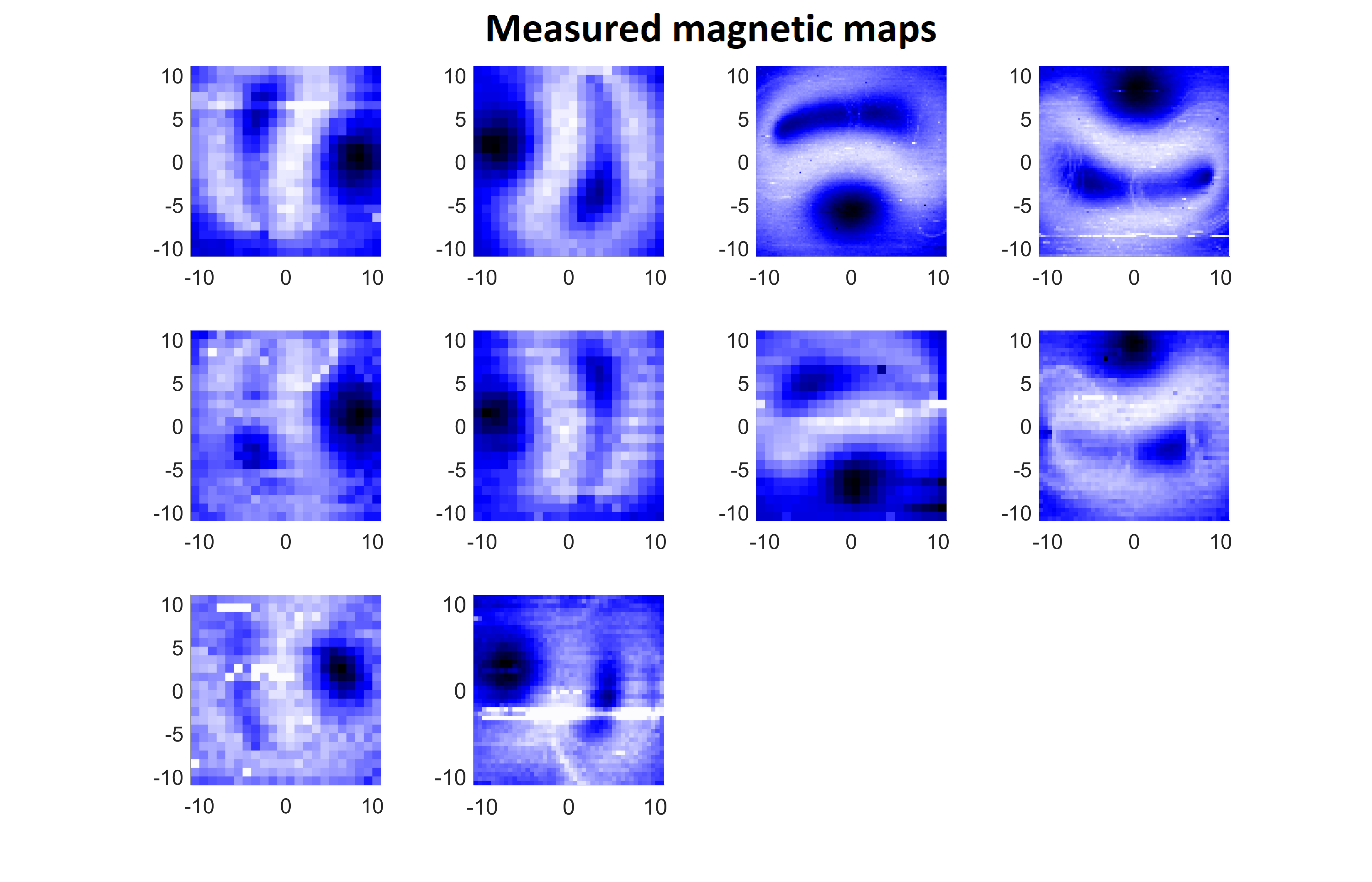}
     \caption{The simulated (top) and measured (bottom) magnetic maps for ST2 are shown. The axes correspond to the magnet position in millimeters, with brightness represented by the color scale. The missing maps have not been observed so far.}
\end{figure*}
\section*{SUPPLEMENTARY NOTE 10: Intersystem crossing rates $\gamma_X$ and $\gamma_Y$}
\begin{figure}
     \centering
         \includegraphics[width=0.45\textwidth]{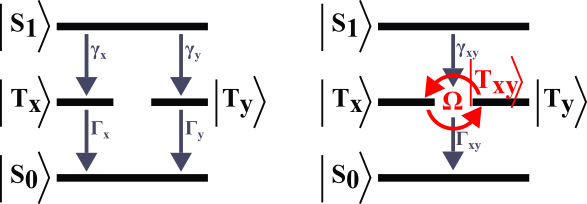}
     \caption{Different decay scenarios within the metastable triplet state with and without an applied microwave (MW) field are considered. Without MW (left), the states are completely disconnected. With MW (right), T$_X$ and T$_Y$ are treated as a combined state.}
\end{figure}
Assessing the rates into the metastable triplet states $\gamma_x$, $\gamma_y$ and $\gamma_z$,  as mentioned in the main text, is not straightforward. 
Currently, these values are treated as fit parameters to match the overall ODMR contrast observed in measurements.
However, some degrees of freedom remains, meaning that multiple combinations of different rates could explain the data.
This multitude of solutions can be reduced by  considering an interesting relation.
A particularly intriguing observation was that the ODMR transition between 
T$_x$ and T$_y$ could never be observed. 
Given the apparent difference in lifetimes, this is rather surprising. 
Explanations such as unfavorable microwave orientations can be ruled out by applying two microwave sources at the same time, which immediately make this ODMR transition visible.
The best alternative explanation we can provide, is a relationship between the populations of the different sublevels.

Generally, the argument for ODMR resonance lines in the electronic level structure for ST2 involves the shift of population from a long-lived state to a short-lived one.
However, it should be apparent that, simultaneously, population from the short-lived state is shifted to the long-lived state. 
In the general case, it is nontrivial that this leads to an effective reduction of the average population in both states combined. 
When comparing the overall lifetime of the metastable triplet, two distinct conditions can be found for the effective lifetime to be independent of the application of driving microwaves. 

The condition for no OMDR line to show up is that the average time the system spends going through the metastable triplet state T, summing over all pathways, is independent from the application of a microwave.
As the rates into the metastable triplet state are for sure independend of any microwaves which are only applied thereafter, this is equivalent to asking that the total population inside the metastable triplet state $n_T$ is the same.

In the situation of interest here, the state T$_Z$ is always independent of T$_X$ and T$_Y$, and its contribution will always bin independent and additive. It can therefore be neglected in the further consideration.

We only need to compare the two situations displayed in Supplementary Figure 3. On the left the two pathways through T$_X$ and T$_Y$ are independent and add up.
On the right however the population goes into a combined state T$_X$ + T$_Y$ from which it decays with one combined Rate $\Gamma_{xy}$.
How do these two situations compare?
On the left we have in the steady state solution.
\begin{equation}
\begin{split}
\dot{n}_{T_x}=0=n_e\gamma_x-n_{T_x}\Gamma_x\\
\dot{n}_{T_y}=0=n_e\gamma_y-n_{T_x}\Gamma_y
\end{split}
\end{equation}
This leads to
\begin{equation}
n_T=n_e \cdot \left(\frac{\gamma_x}{\Gamma_x}+\frac{\gamma_y}{\Gamma_y}\right)
\end{equation}
In the situtation on the right in Figure S3, we need to consider the population inside the combined state Tx+Ty. The rates into this state is simply the addition of the two rates: $\gamma_{xy}=\gamma_x+\gamma_y$.
The rate out of this combined state however is given by $\Gamma_{xy}=\frac{\Gamma_x+\Gamma_y}{2}$.
If we denote the right siutation as the primed on this leads to
\begin{equation}
n_T'=n_e \cdot \left(\frac{\gamma_{xy}}{\Gamma_{xy}}\right)=n_e \cdot \left(\frac{\gamma_{x}+\gamma_{y}}{\frac{\Gamma_{x}+\Gamma_{y}}{2}}\right)
\end{equation}
To find all situations in which the application of a microwave has no effect on the system we simply need to find all solutions to the equation
$n_T=n_T'$.

This simple but rather lengthy calculation leads to two conditions.
The first condition is the obvious equality of the two lifetimes. The second, and more interesting, condition is:
\(\gamma_x / \Gamma_x = \gamma_y / \Gamma_y\).
Given the clear difference in lifetime between the T$_x$ and T$_y$ states, we first assumed that this condition is fulfilled.
However, as mentioned in the main manuscript, directly assessing the rates into the metastable triplet states $\gamma_x$,$\gamma_y$ and $\gamma_z$  is not straightforward. 
It can only be assessed indirectly through simulations that reproduce the ODMR contrast of the resonance lines. This comparison clearly contradicts the above relation and instead leads to $\gamma_x=\gamma_y$, which is required to match the difference in ODMR contrast observed in Figure 2c of the main manuscript.
\section*{SUPPLEMENTARY NOTE 11: Cross relaxation via nearby electron spins: Brightness rings}
An intriguing feature observed in most ST2 centers, which has not been addressed until now, is the presence of ring-like structures in the magnetic maps, as for example in Figure 3a of the main manuscript.
In the provided example, these rings are relatively subtle and not particularly pronounced.
For a clearer demonstration of this phenomenon, we have included a more striking example in Figure S4.
\begin{figure}
     \centering
         \includegraphics[width=0.48\textwidth]{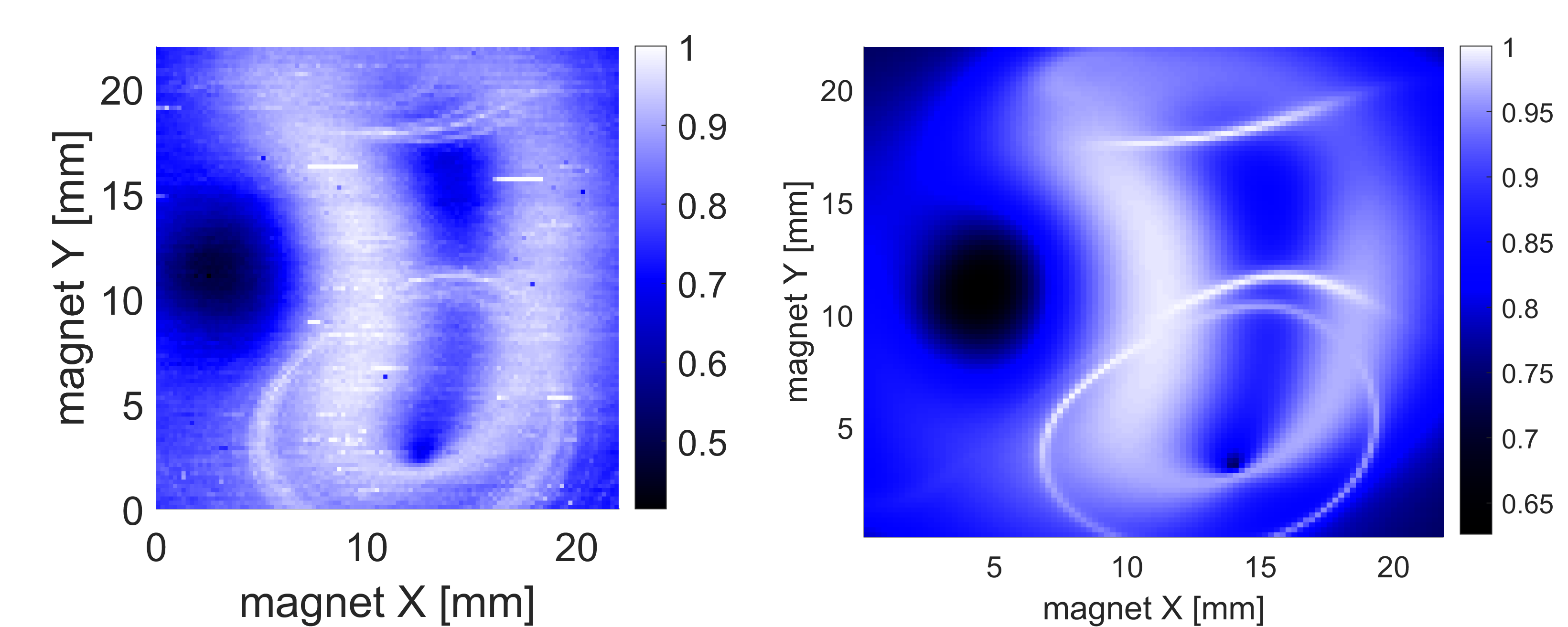}
     
     \caption{The left side displays a magnetic map of an ST2 center exhibiting relatively strong additional rings. These rings arise due to cross-relaxation processes involving nearby electron spins. On the right, a simulation result is presented that incorporates the effects of cross relaxation.}
\end{figure}
To understand the appearance of this phenomenon, it is useful to recall that the creation process of ST2 centers inevitably introduces high lattice damage within the diamond. 
This damage significantly increases the likelihood of nearby defects, which explains why the majority of, rather than just a select few, ST2 centers exhibit at least weak effects of cross relaxation.
One common defect is a trapped electron with spin 1/2. 
These trapped electrons, situated near the ST2 center, experience a similar magnetic field, making theoretical modeling straightforward. 
The external magnetic field splits the levels of the spin-1/2  system, with transitions occurring at the Larmor frequency.
Cross relaxation happens when the Larmor frequency matches a transition within the ST2 center's metastable triplet. This results in spin flips for both the electron and the triplet, effectively mimicking microwave-driven transitions.
This renders the simulation equivalent to the one described in Supplementary Note 9.
If cross relaxation, links states with different lifetimes, the population of the metastable triplet generally decreases, causing the center to emit more light. 
This is visible as the bright rings in Figure S4, shown in measurements on the left and simulations on the right.

\bibliography{LiteraturSup}
%\begin{thebibliography}{34}%

%\end{thebibliography}%